\documentclass{osa-article}
\usepackage[utf8]{inputenc} 
\usepackage[T1]{fontenc}
\usepackage{graphicx}
\usepackage{grffile}
\usepackage{longtable}
\usepackage{wrapfig}
\usepackage{rotating}
\usepackage[normalem]{ulem}
\usepackage{amsmath}
\usepackage{textcomp}
\usepackage{amssymb}
\usepackage{capt-of}
\usepackage{hyperref}
\journal{osac}
\usepackage{subcaption}
\usepackage{float}
\usepackage{xcolor}

\let\origeqref\eqref
\newcommand{\eqrefn}[1]{Eq. \origeqref{#1}}
\newcommand{\neff}{n_{\text{eff}}}

\newcommand{\intd}{\,\textup{d}}

\newcommand{\lcwl}{\lambda_{\text{cwl}}}

\newcommand{\cone}{\theta_{\text{cone}}}
\newcommand{\cra}{\theta_{\text{CRA}}}

\newcommand{\phiapproxmin}{\neff\sqrt{-\dfrac{2\lambda}{\lcwl}}}
\usepackage{epstopdf}
\usepackage{tikz}
\date{\today}
\hypersetup{
 pdfauthor={Thomas Goossens}, 
 pdftitle={},
 pdfkeywords={}, 
 pdfsubject={},
 pdfcreator={Emacs 26.1 (Org mode 9.1.14)}, 
 pdflang={English}}
\begin{document}

\title{Vignetted-aperture correction for spectral cameras with integrated thin-film Fabry-Pérot filters}

\author{Thomas Goossens\authormark{1,2,*}, Bert Geelen\authormark{2}, Andy Lambrechts\authormark{2}, Chris Van Hoof\authormark{2,1}}

\address{\authormark{1}Department of Electrical Engineering, KU Leuven, Leuven 3001, Belgium\\
\authormark{2}imec vzw, Kapeldreef 75, 3001 Leuven, Belgium\\
\email{\authormark{*}contact@thomasgoossens.be\\ \url{https://orcid.org/0000-0001-7589-5038}}}

\begin{abstract*}
Spectral cameras with integrated thin-film Fabry-Pérot filters
have become increasingly important in many applications. These
applications often require the detection of spectral features at
specific wavelengths or to quantify small variations in the spectrum.
This can be challenging since thin-film filters are sensitive to the angle of
incidence of the light. 
In prior work we modeled and corrected for the distribution of incident angles for an ideal finite
aperture. Many real lenses however experience vignetting. Therefore in this article we generalize our model to the more common case of a vignetted aperture, which changes the distribution of incident angles.   
We propose a practical method to estimate the model parameters and
correct undesired shifts in measured spectra. This is experimentally
validated for a lens mounted on a visible to near-infrared spectral camera. \textit{© 2019 Optical Society of America.} \\

\href{https://doi.org/10.1364/AO.58.001789}{https://doi.org/10.1364/AO.58.001789}
\end{abstract*}

\section{Introduction}
\label{sec:org24383ff}

Spectral imaging is a technique that combines photography and
spectroscopy to obtain a spatial image of a scene and for each point
in that scene and also sample the electromagnetic spectrum at many wavelengths. This spectrum can be
used as a 'fingerprint' to identify different materials in the scene.

In some applications, small differences in the spectrum have to be
quantified. Spectral imaging is required if these differences are too
small to be detected with RGB color cameras. Other applications require the
detection of spectral features at specific wavelengths \cite{Shippert2004}.
Therefore, spectral imaging has the potential to increase selectivity
in machine learning applications. Also, having a spectrum enables physical
interpretation of the acquired spectral data (e.g. for material identification) \cite{Lu2014,Gowen2007}. 

In recent years spectral cameras have been developed with
integrated thin-film Fabry-Pérot filters on each pixel of an image
sensor \cite{Geelen2014,Tack2012}. To obtain an image, the sensor array is placed in the image
plane of an objective lens which focuses light from the scene onto each pixel.

Focused light from a finite aperture widens and shifts the
transmittance peak of thin-film Fabry-Pérot filters because these
filters are angle-dependent \cite{Goossens2018}. This angle-dependent property causes undesired shifts in the
measured spectra which limit the accuracy of the spectral imaging
technology. Therefore correcting these shifts is important for improving
the performance of many applications including machine learning. 

In previous work we proposed a model that was used to successfully
correct the undesired shifts in measured
spectra \cite{Goossens2018}. However, this model is only valid for a lens that
shows no significant optical vignetting. 

In this article we generalize our original model to allow for an aperture with
optical vignetting. We also propose a practical algorithm
to estimate the model parameters. The results are therefore of great practical importance since optical vignetting
is a common phenomenon. Even more so for very compact and less
expensive lenses. Hence this paper enables the use of such lenses and
enables spectral imaging for more price sensitive applications.

\section{Theory and methods}
\label{sec:org5cb40c6}
In this section, we discuss how a spectral camera can be modeled
(Subsection \ref{org8a59309}) and how the effect of an ideal finite aperture was modeled in earlier work
(Subsection \ref{org7788f33}). We extend the model to include
optical vignetting (Subsection \ref{orgcda1569} and
\ref{orgbc32e67}) and how to correct for it (Subsection
\ref{org9f2472b}). We conclude with an algorithm to estimate the model
parameters (Subsection \ref{org5b1dfa1}).

\subsection{Spectral imaging model}
\label{sec:org9e981b2}
\label{org8a59309}

\begin{figure}[htbp]
\centering
		 \includegraphics[width=0.6\linewidth]{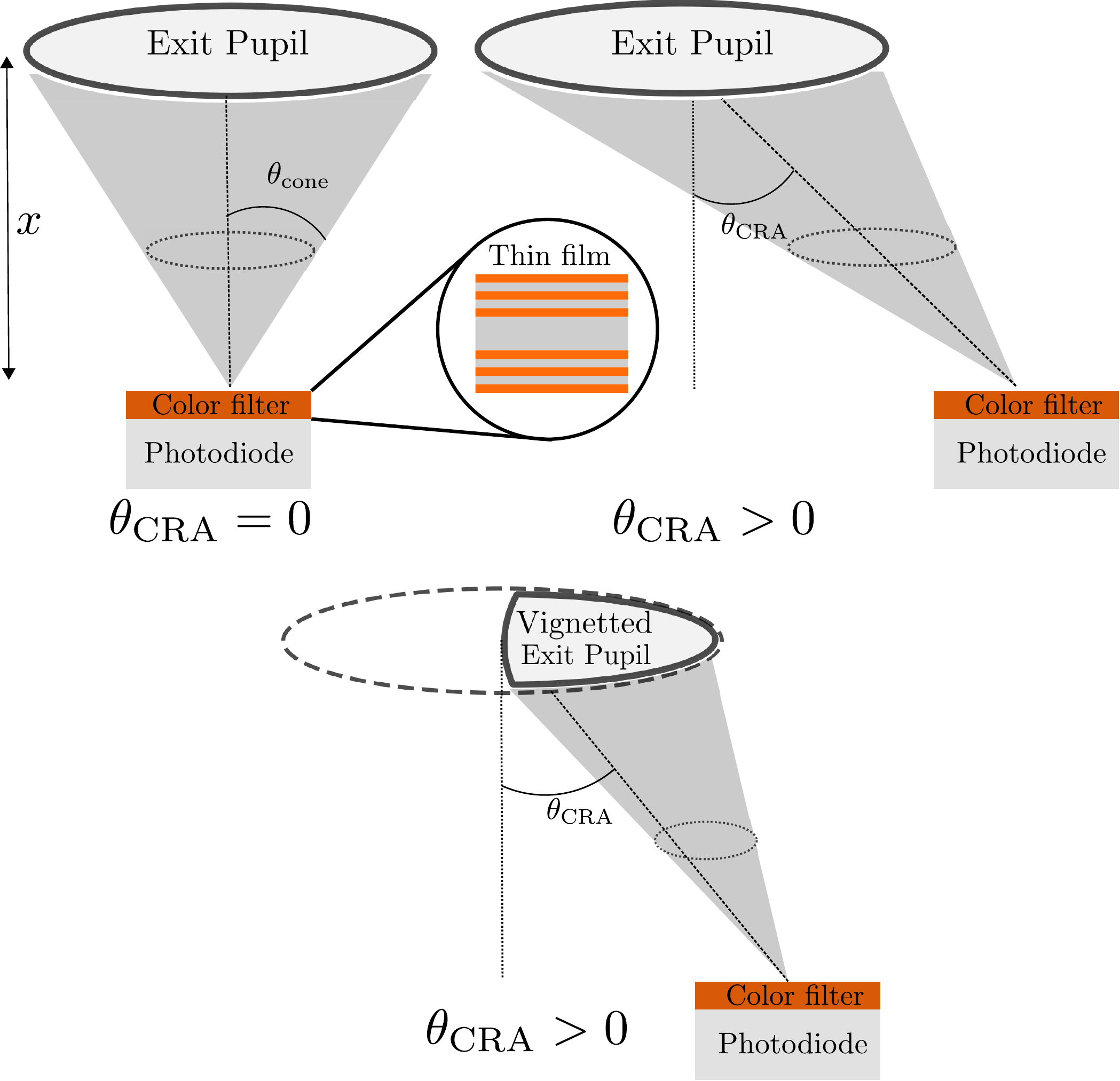}
		 \caption{\label{fig:cameramodel} A schematic representation of how an aperture focuses light on the spectral imaging sensor with integrated thin-film optical filters. In the case of vignetting, part of the exit pupil is cut off. Only the vignetted exit pupil then contributes to focusing the light.}
\end{figure}

The imaging system used in this article consists of an objective lens with a finite aperture (exit pupil)
which focuses the light onto pixels of an imaging sensor with
integrated thin-film interference filters (Fig. \ref{fig:cameramodel}). 
The output \(\text{DN}\) (Digital Number) of a pixel with integrated filters can be modeled as

\begin{equation}
\label{eq:pixel}
\text{DN} = \int_{\lambda_\mathrm{l}}^{\lambda_\mathrm{r}} T(\lambda) \cdot L(\lambda) \intd \lambda.
\end{equation}

Here \(L(\lambda)\) is the irradiance spectrum of the light incident at the pixel and \(T(\lambda)\) is the
transmittance of the filter measured under orthogonal collimated
light conditions. The gain factors are assumed to be equal to 1 and are
therefore omitted \cite{janesick2007photon}. The limits of the
integral describe the bandpass range of the spectral camera.

For a given f-number (characterized by half-cone angle \(\cone\)) and chief ray angle \(\cra\), the effect of the
aperture and optical vignetting on the filter with central wavelength \(\lcwl\) is modeled as a convolution of
\(T(\lambda)\) with a kernel \(K_{\cone,\,\cra}(\lambda)\) such  that 
\begin{equation}
\label{eq:pixel-kernel}
 \text{DN} = \int_{\lambda_\mathrm{l}}^{\lambda_\mathrm{r}} (K_{\cone,\,\cra} * T) (\lambda) \cdot L(\lambda) \intd \lambda,
 \end{equation}
with \(*\) being the convolution operator and the f-number being the
ratio of the effective focal length to the diameter of the entrance
pupil. 

Below in Table \ref{tab:orga607052}, the most important symbols are summarized.
\renewcommand{\arraystretch}{1.2}
\begin{table}[htbp]
\caption{\label{tab:orga607052}
List of the main symbols and their meaning.}
\centering
\begin{tabular}{lll}
Symbol & unit & Meaning\\
\hline
T(\(\lambda\)) &  & Transmittance of a filter under\\
 &  & orthogonal collimated conditions\\
\(\lcwl\) & nm & central wavelength of filter\\
\(f_{\#}\) &  & f-number\\
\(f_{\#,W}\) &  & working f-number\\
\(x\) & mm & distance from pixel to exit pupil plane\\
\(d\) & mm & distance from optical axis\\
\(\cra\) & rad & chief ray angle\\
\(\cone\) & rad & half-cone angle of the image side\\
 &  & light cone\\
\(R\) & mm & Radius of the circular exit pupil\\
\(P\) & mm & Radius of the projected vignetting circle\\
\(h\) & mm & Distance of exit pupil to entrance circle\\
\end{tabular}
\end{table}

\subsection{Finite aperture correction}
\label{sec:org61d4bd5}
\label{org7788f33}

In spectral camera designs where interference filters are spatially arranged
behind an objective lens, understanding the impact of focused
light from a finite aperture becomes essential.
In this section we discuss the effect of the angle of incidence on
thin-film Fabry-Pérot filters and the main results of prior work on the effect of
focused light. 

Thin-film Fabry-Pérot filters are constructed by combining multiple
layers of a high and
a low refractive index material \cite{macleod2001thin}. The filters
approximate behavior of the ideal Fabry-Pérot etalon.

The transmittance characteristics of interference filters change with
the angle of incidence of the light. The larger the angle of
incidence, the more the transmittance peak shifts towards shorter wavelengths.

The angle-dependent behavior of multilayer thin-film filters can be simulated using the
transfer-matrix method \cite{macleod2001thin}. However, for angles up
to 40 degrees it was shown that the filter can be modeled
as an ideal Fabry-Pérot etalon \cite{Pidgeon1964,macleod2001thin}. This idealized etalon has an effective cavity material with an effective refractive index
\(\neff\) which depends on the materials used.

The shift in central wavelength of a thin-film Fabry-Pérot filter with
effective refractive index \(\neff\) for an incident angle \(\phi\) is
well approximated by \(\Delta(\phi)\) which is defined as 
\begin{equation}
\label{eq:tiltshift}
\Delta(\phi) = -\lcwl \left(1-\sqrt{1-\dfrac{\sin^2\phi}{\neff^2}}\right),
\end{equation}
with \(\lcwl\) the central wavelength of the transmittance
peak. We have changed the sign convention compared to
Eq. (3) in \cite{Goossens2018}. It is natural to consider a negative
shift because the central wavelength becomes shorter.

This shift in central wavelength can also be modeled as a convolution
of the initial transmittance \(T(\lambda)\) with a shifted Dirac distribution such that
\begin{equation}
\label{eq:tiltkernel}
T_\phi(\lambda) \sim T(\lambda) * \delta\left(\lambda -  \Delta(\phi) \right).
\end{equation}

To model the effect of a finite aperture, focused
light can be interpreted as a distribution of incident angles characterized by
two parameters: the cone angle \(\cone\) (or f-number) and chief ray angle
\(\cra\) (Fig. \ref{fig:cameramodel}). 
The analysis can be simplified by decomposing the focused beam of light in
contributions of equal angles of incidence.
In earlier work we have shown that the effect of the aperture
can be modeled by a convolution of the transmittance of the filter
with a kernel \(K_{\cone,\,\cra}(\lambda)\) \cite{Goossens2018}. The kernel was defined as 
\begin{equation}
\label{eq:asymptoticold}
K_{\cone,\,\cra}(\lambda) \sim \dfrac{2\,\neff^2}{\lcwl} \cdot \dfrac{\displaystyle \eta\left(\phiapproxmin \right) } {\pi \tan^2 \cone},
\end{equation}
as used in \eqrefn{eq:pixel-kernel} and with \(\eta\) as it will be defined in \eqrefn{eq:eta}.
The changes in the transmittance of the filters cause undesired
shifts in the measured spectra. 

Two differences to the corresponding
Eq. (15) in \cite{Goossens2018} must be pointed out. First, because of the change in sign
convention (see \eqrefn{eq:tiltshift}), there is a sign difference
under the square root sign. Second,
in this equation, the symbol \(\eta\) is equal to \(\gamma\) as used in
Eq. (15) in \cite{Goossens2018}, where it was used for the case
without vignetting. In this article, the symbol \(\gamma\) will be
reserved for the more general case of vignetting.

In prior work we showed that the mean
value \(\bar{\lambda}\) of the kernel 
can quantify the shift in central wavelength. The mean value of the
kernel was asymptotically equivalent to
\begin{equation}
\label{eq:idealmean}
 \bar{\lambda} \sim -\lcwl \left(\dfrac{\cone^2}{4\neff^2} +
 \dfrac{\cra^2}{2\neff^2}\right),\ \text{for } \cone,\cra \rightarrow 0.
\end{equation}

This formula can be used to correct the measured spectra. 
The wavelength at which the response of each pixel is plotted is
updated as
\begin{equation}
\label{eq:idealcorrection}
 \lcwl^{\text{new}} = \lcwl + \bar{\lambda} =\lcwl \left(1 - \dfrac{\cone^2}{4\neff^2} -
 \dfrac{\cra^2}{2\neff^2}\right).
\end{equation}

In this article an important extension of the above analysis is
presented. The model is generalized to include optical vignetting
which is a common phenomenon in many real lenses.

\subsection{Vignetting}
\label{sec:orgccf97ab}

Vignetting is the fall-off in intensity that is observed in an image
that was taken of a uniformly illuminated scene. There are many types of
vignetting that can occur simultaneously. Examples are the
cosine-fourth fall-off, optical vignetting, mechanical vignetting,
pixel vignetting and pupil aberrations. 

The 'cosine-fourth' fall-off, also called natural vignetting is an
effect intrinsic to even ideal thin lenses. It is caused by
projected areas of the lens and pixels and the inverse square law. It
implies that, under
certain assumptions, the intensity of the light will be scaled by
a factor \(\cos^4 \cra\) \cite{Gardner1947}. In wide-angle lenses the
effect is more pronounced and is often corrected for by design \cite{Asada1996,Smith2000}. 

\begin{figure}[H]
\centering
		 \includegraphics[width=0.8\linewidth]{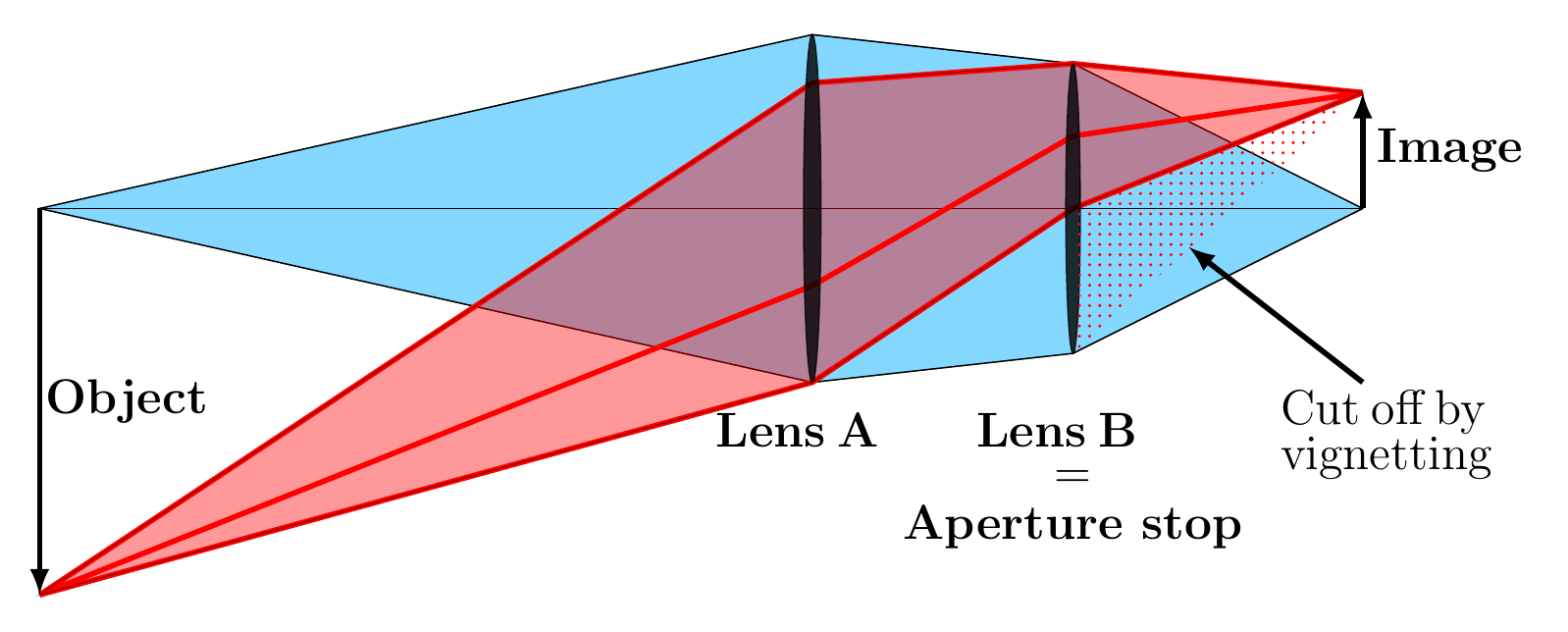}	
		 \caption{\label{fig:vignetting} \textbf{Illustration of optical vignetting.} The off-axis point does not receive light from the complete aperture stop (lens B). This optical vignetting is caused by the limited diameter of lens A, causing lens B to be partially shaded \cite{Smith2000}. } 
\end{figure}

Optical vignetting occurs due to the physical length of a lens system or the
position of a limiting aperture somewhere in the optical path \cite{Smith2000,ray2002applied}. In
essence, lens elements can shade other lens elements
(Fig. \ref{fig:vignetting}). This causes
part of the light beam to be cut off. The amount of optical vignetting
depends on the aperture size and is less pronounced for smaller apertures
(high f-numbers).

In mechanical vignetting part of the scene is externally obstructed by for
example the lens hood. This can cause the entrance pupil to be
partially shaded \cite{ray2002applied}. In the proposed model there will be no fundamental difference with
optical vignetting. 

Pixel vignetting is an effect in CMOS digital cameras where the
photodiode is positioned at the end of a tunnel in the back end of
line of the image sensor \cite{Catrysse2002}.
Light can only reach the photodiode via the tunnel which when illuminated at oblique incidence,
casts a shadow on the photodiode. Pixel vignetting behaves
similar to optical vignetting since it also cuts of part of the light
beam. 

Pupil aberrations describe how the aperture is not uniformly
illuminated \cite{Aggarwal2001}. Other types of aberrations cause the exit pupil to
move or change in size \cite{Hazra2013,Sasian2006}.

In the remainder of this article it is assumed that optical and mechanical
vignetting are the dominant effects. Unless when specified, we will use the
term 'vignetting' and 'vignetted aperture' to imply optical and mechanical vignetting. 

\subsection{Modeling a vignetted aperture}
\label{sec:org7e1c922}
\label{orgcda1569}

A vignetted aperture is an aperture that is partially cut off as a
result of optical or mechanical vignetting. This aperture will be seen
as a differently shaped exit pupil for each pixel (Fig. \ref{fig:cameramodel}). In this article the 'exit pupil' is the image
of the aperture stop (full circle) and the 'vignetted exit pupil' is the
part of the exit pupil still focuses light onto the pixel.

\begin{figure}[H]
\centering
		 \includegraphics[width=0.8\linewidth]{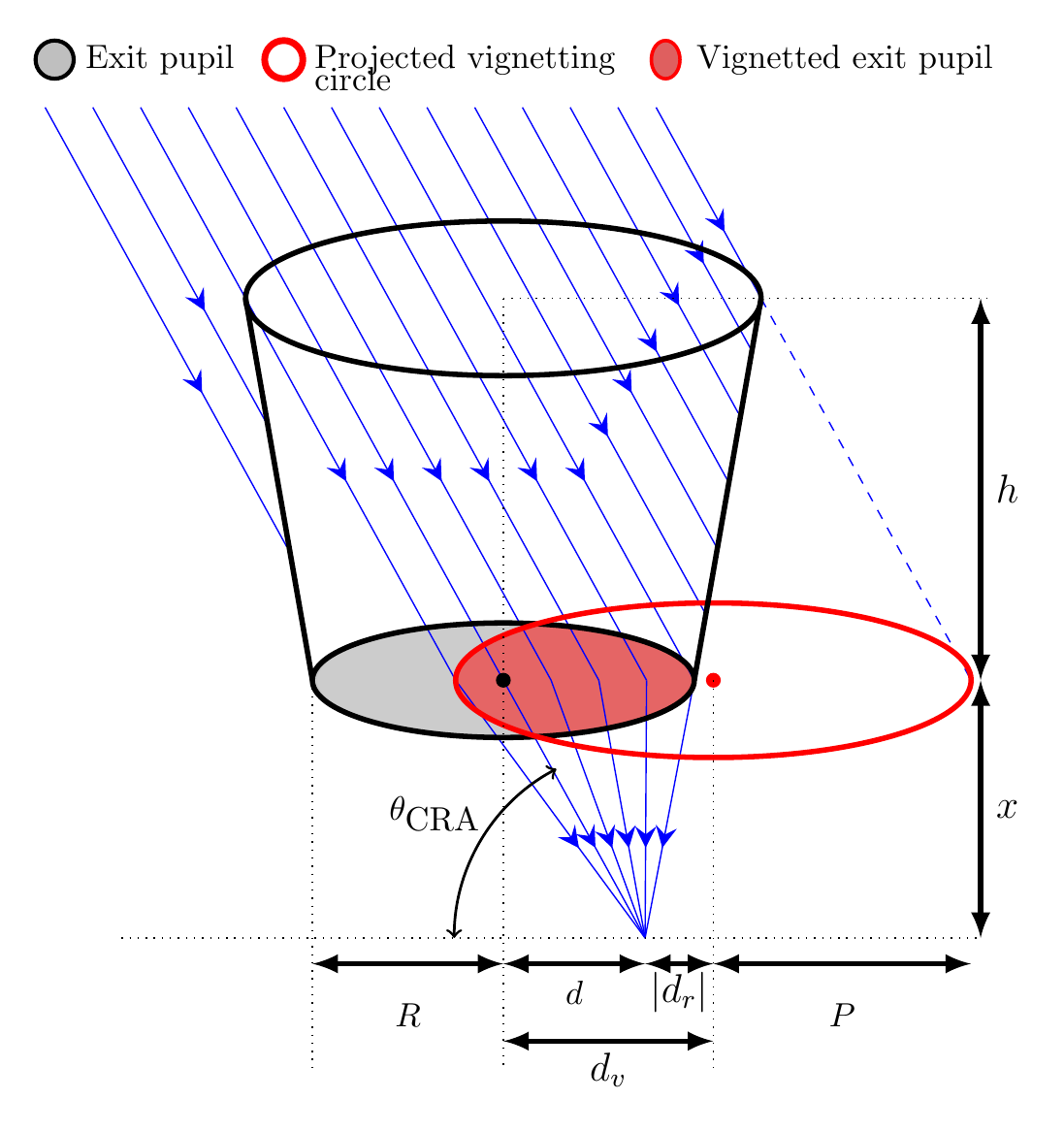}	
		 \caption{\label{fig:variablecone} Optical vignetting is modeled with a finite aperture (exit pupil) encapsulated in a tube illuminated with collimated light. The tube has an opening of radius $P$ at a distance $h$.
		          The vignetted exit pupil is the overlapping area between the exit pupil and the projected vignetting circle. Only the vignetted exit pupil still focuses light.}
\end{figure}

An insightful approach to understand the shape of a vignetted aperture is provided by the
variable cone model \cite{Asada1996}. This model assumes that there is
one finite aperture (exit pupil) that is encapsulated in a tube with
length \(h\) and an entrance circle of radius \(P\)
(Fig. \ref{fig:variablecone}). It also assumes that the incident light
is collimated. 

The exit pupil is the image of the limiting aperture (aperture stop) somewhere in the lens system. The second most limiting
aperture will be responsible for most of the vignetting. The image of this second
aperture is modeled by the entrance circle of radius \(P\) of the tube
in Fig. \ref{fig:variablecone}. The projection of this entrance circle
onto the plane of the exit pupil is called the projected vignetting circle.

Because of its physical dimensions, the tube casts a shadow onto the
exit pupil plane.  The part of the exit pupil that still contributes to
focusing the light is then the intersection of the exit pupil and the
projected vignetting circle (red area in
Fig. \ref{fig:variablecone}). This remaining part we will call the
'vignetted exit pupil'.

The assumption of collimated light
simplifies the analysis. In real lens systems however, the light might
not be collimated. Yet, even in these cases the exit pupil is still cut off by some
circle \cite{Smith2000}. The assumption will therefore only impact the
fitted values for \(P\) and \(h\).

To calculate the shape of the vignetted exit pupil, the distances \(d\)
and \(d_v\) need to be defined (Fig. \ref{fig:variablecone}).
The position of a pixel \(d\) is characterized by the distance to the
optical axis such that 
\begin{equation}
\label{eq:d}
d = x \tan\cra.
\end{equation} 
The position \(d_v\) of the center of the projected vignetting circle
\begin{equation}
\label{eq:dv}
d_v = h \tan\cra.
\end{equation}
From \eqrefn{eq:d} and \eqrefn{eq:dv} it follows that the
relative size of \(h\) and \(x\) causes qualitative differences. This
is discussed in the next section.

The variable cone model offers useful intuition about the behavior of optical vignetting.
The larger the radius \(P\) relative to \(R\), the larger \(d_v\) (and hence
\(\cra\)) needs to become before part of the exit pupil is cut off. This
intuitively explains why for smaller apertures (small R), there is
less optical vignetting. 

The variable cone model has been used before to model the intensity
fall off. However, it has never been used to calculate the
distribution of incident angles. 
Therefore, in the next section, we calculate how the shape of the vignetted exit
pupil changes the distribution of incident angles for each pixel. We then
calculate how this distribution impacts the transmittance of the
integrated thin-film Fabry-Pérot filters. 

\subsection{Focused light from a vignetted aperture}
\label{sec:org40919ff}
\label{orgbc32e67}

Each filter has a transmittance \(T(\lambda)\) when illuminated under
orthogonal, collimated conditions. When used with a lens, and because
of vignetting, each filter
on the sensor array sees a different distribution of incident angles. 
The resulting transmittance is defined as \(\hat{T}_{\cone,\,\cra}(\lambda)\).

\begin{figure}[htpb!]
\centering
		 \includegraphics[width=0.7\linewidth]{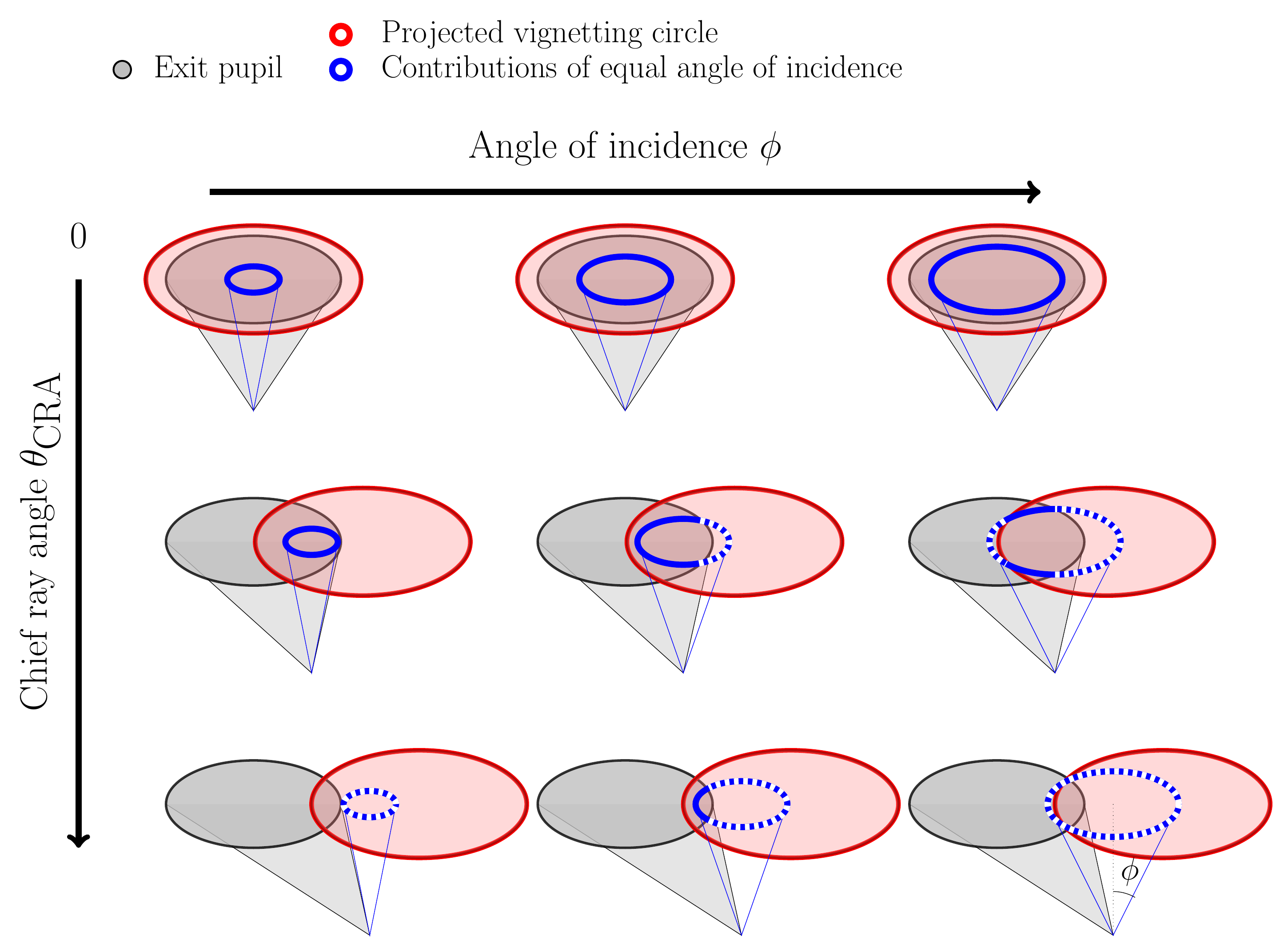}
		 \caption{\label{fig:grid} For each $\cra$, the light beam is decomposed in contributions with
		  equal angles of incidence. The weight of a contribution is measured by the length of the blue arc
		   within the aperture. Parts of the blue arc can be cut off by the projected vignetting circle.} 
\end{figure}

To analyze the focused light beam from the vignetted aperture, the
light beam can be decomposed in contributions of equal angle of
incidence \(\phi\) (Fig. \ref{fig:grid}). Each contribution can then be treated
separately using the tilt angle model from \eqrefn{eq:tiltkernel}.

The resulting transmittance is a linear combination of contributions
of the form of \eqrefn{eq:tiltkernel}, each representing 
a different angle of incidence. The weight of each
contribution is the infinitesimal area \(\textup{d}A\) of the aperture
that contributes to the same angle of incidence \(\phi\) (marked by blue
in Fig. \ref{fig:grid}). The transmittance thus becomes
\begin{equation}
\label{eq:effectaperture}
\hat{T}_{\cone,\,\cra}(\lambda) = 
\dfrac{\displaystyle \iint\limits_{\tiny \substack{\text{Vignetted} \\ \text{exit pupil}}} T_\phi(\lambda) \intd A}{\displaystyle \iint\limits_{\tiny \substack{\text{Vignetted} \\ \text{exit pupil}}} \intd A},
\end{equation}
where the normalization is required because of conservation of energy.

Because of linearity of the operators, \(\hat{T}_{\cone,\,\cra}(\lambda)\) can be written
as a convolution of the transmittance \(T(\lambda)\) at orthogonal
collimated conditions with a kernel \(K_{\cone,\,\cra}(\lambda)\) such that
\begin{eqnarray}
\hat{T}_{\cone,\,\cra}(\lambda)
&=& T(\lambda)* \dfrac{\displaystyle \iint\limits_{\tiny \substack{\text{Vignetted} \\ \text{exit pupil}}} \delta(\lambda - \Delta(\phi)) \intd A}{\displaystyle \iint\limits_{\tiny \substack{\text{Vignetted} \\ \text{exit pupil}}} \intd A} \label{eq:kernelintegral}\\
&=& T(\lambda)* \dfrac{\displaystyle \iint\limits_{\tiny \substack{\text{Vignetted} \\ \text{exit pupil}}} \delta(\lambda - \Delta(\phi)) \intd A}{A(\cone,\cra)} \label{eq:kernelintegral}\\
&=& T(\lambda)* K_{\cone,\,\cra}(\lambda).
\end{eqnarray}
 Here \(A(\cone,\cra)\) is the area
of the exit pupil which, because of vignetting, also changes with the
chief ray angle (see Appendix \ref{org3f6ca83}).

\begin{figure}[h!]
\centering
	 \begin{subfigure}{0.7\linewidth} 
		 \includegraphics[width=0.8\linewidth]{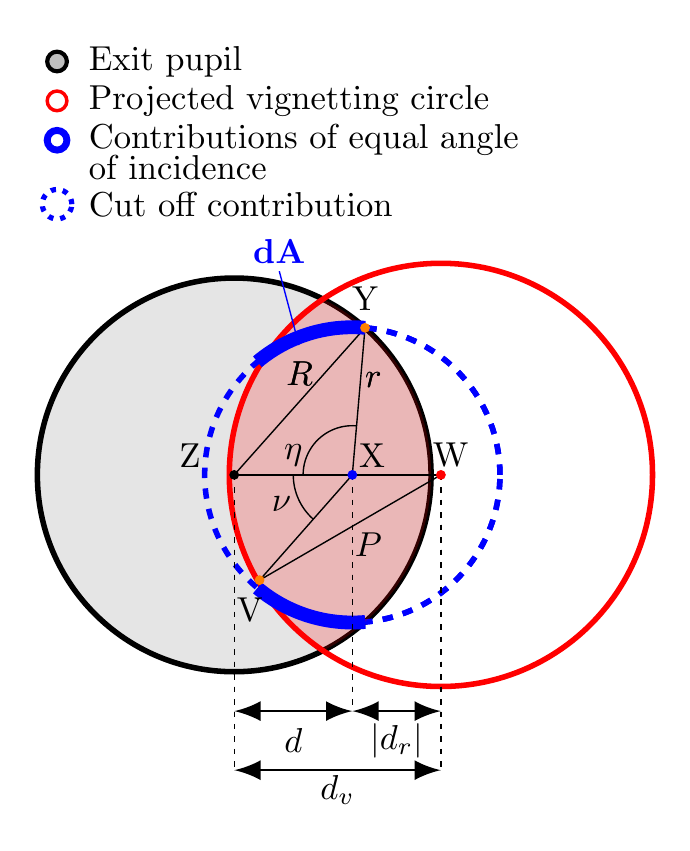}
		 \caption{Top view of the exit pupil plane for $h\ge x$ or equivalently $d_v \ge d$. A part of the contributions of equal angle of incidence is cut off so that only the thick blue line remains. \label{fig:decomp-top}} 
	 \end{subfigure}
	 \begin{subfigure}{0.7\linewidth} 
		 \includegraphics[width=0.8\linewidth]{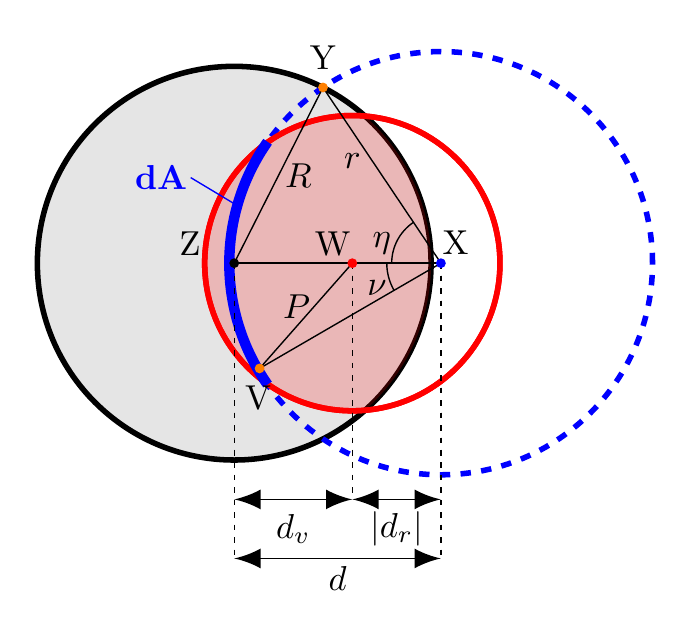}
		 \caption{Top view of the exit pupil plane for $h<x$ or equivalently $d_v < d$. The part cut off by the vignetting circle is qualitatively different than in Fig. \ref{fig:decomp-top}. \label{fig:decomp-top-case2}} 
	 \end{subfigure}
	 \caption{  \label{fig:decomp} Decomposition of the light cone from the aperture in contributions of equal angle of incidence $\phi$. The weight of each contribution is the infinitesimal area $\intd A$ (blue). Here $d$ is the distance to the pixel from the optical axis.}
\end{figure}

The set of all possible rays that have the same incident angle on a
pixel is a hollow cone with half-cone angle \(\phi\) 
(the blue cone in Fig. \ref{fig:grid}). The intersection of this cone with
the exit pupil plane is a circle with radius \(r =x\tan\phi\).

In reality there are only rays coming from within the exit pupil. This
subset is the part of the blue circle that lies within
the exit pupil (Fig. \ref{fig:grid}). Its contribution is
calculated as the infinitesimal area
\(\textup{d}A\) of a ring segment (Fig. \ref{fig:decomp-top}):
\begin{equation}
\textup{d}A = 2\gamma(\arctan \dfrac{r}{x})r\textup{d}r = 2\gamma(\phi)r\textup{d}r,
\end{equation}
with \(\gamma(\phi)\) being the angle of the arc within the
exit pupil that contributes to the same angle of incidence \(\phi =
\arctan \frac{r}{x}\).

The angle \(\gamma\) will be defined as a function of two other angles:
\(\eta\) and \(\nu\).
The angle \(\eta\) is the angle that describes the contribution 
in the absence of vignetting (Fig. \ref{fig:decomp-top}). It is only limited by
the area of the exit pupil. The effect of vignetting will be that only a subsection
of the arc described by \(\eta\) will be relevant. This relevant part is
calculated using the angle \(\nu\). It is the angle that describes what
part of the arc is cut off (Fig. \ref{fig:decomp-top}) or kept (Fig. \ref{fig:decomp-top-case2}) by the vignetting circle. 

Taking into account that \(r=x\tan\phi\), the angle \(\eta(\phi)\) is determined by the law of cosines in
\(\Delta\text{XYZ}\) (Fig. \ref{fig:decomp-top}) as
 \begin{equation}
\label{eq:eta}
 \eta(\phi) = \text{Re}\left( \textup{arccos} \dfrac{d^2-R^2+r^2}{2dr}\right).
 \end{equation}
By taking the real part of the inverse cosine, the case in which there
is a complete contributing circle 
within the aperture is also modeled (Fig. \ref{fig:grid})
\cite{Goossens2018}. This is
because \(\text{Re} (\arccos z) = \pi,\, \text{for}\, z\leq-1\).  

Similarly, \(\nu(\phi)\) is defined by applying the law of cosines in
\(\Delta\text{XWV}\) (Fig. \ref{fig:decomp}) such that for \(h\ge x\),
\begin{equation}
\nu(\phi) = \pi - \text{Re}\left( \textup{arccos} \dfrac{d_r^2-P^2+r^2}{2 |d_r| r}\right).
\end{equation}

By defining \(d_r = d-d_v\), the above expression can be simplified as
\begin{equation}
\label{eq:nu}
\nu(\phi) = \text{Re}\left( \textup{arccos} \dfrac{d_r^2-P^2+r^2}{2d_r r}\right),
\end{equation}
which covers both \(h\ge x\) and \(h<x\).

From these definitions and Fig. \ref{fig:decomp-top} it follows that the resulting contributing arc is defined as
\begin{equation}
\label{eq:gamma}
\gamma(\phi) =
\begin{cases} 
  \max(\eta(\phi)-\nu(\phi),0) & h\ge x\\
  \min(\eta(\phi),\nu(\phi)) & h<x
\end{cases},
\end{equation}
where \(\nu(\phi)\) is the angle that is either cut off or kept, depending
on the relative size of \(h\) and \(x\). In the case \(h\ge x\), \(\nu\)
represents the angle that is cut off by the vignetting circle (Fig. \ref{fig:decomp-top}). 
Therefore it must be subtracted from \(\eta\). The \(\max\) operator is needed because there is either a
positive contribution or no contribution. A negative angle for \(\gamma\) has no physical meaning. 

In the case \(h<x\), \(\nu\) represents the angle of the arc that is not cut off
(Fig. \ref{fig:decomp-top-case2}). This is modeled by taking the mininum of \(\eta\) and \(\nu\).

To work with the angle of incidence \(\phi\), \(r\) is
substituted with \(r = x \tan \phi\) such that
\begin{equation}
\textup{d}A = 2x^2\gamma(\phi) \dfrac{\tan \phi}{\cos^2 \phi} \textup{d}\phi.
\end{equation}
The integral then becomes
\begin{equation}
\label{eq:kernelaftersubstitution}
 K_{\cone,\,\cra}(\lambda) = \dfrac{\displaystyle \int_{\phi_\text{min}}^{\phi_\text{max}}   2x^2 \gamma(\phi) \dfrac{\tan \phi}{\cos^2 \phi} \delta(\lambda - \Delta(\phi)) \intd \phi}{A(\cone,\cra)}.
\end{equation}

Here \(\phi_\text{min}\) and \(\phi_\text{max}\) are the smallest and
largest incident angles coming from the aperture (See Appendix \ref{org9acc08e}).

An asymptotic approximation of the kernel is
\begin{equation}
\label{eq:asymptotic}
K_{\cone,\,\cra}(\lambda) \sim \displaystyle  \frac{2\,\neff^2}{\lcwl} \cdot \dfrac{\gamma\left(\phiapproxmin \right) } {A(\cone,\cra)} x^2,
\end{equation}
as used in \eqrefn{eq:pixel-kernel}. For the derivation see Appendix \ref{org3f6ca83}.

\begin{figure}[H]
\centering
		 \includegraphics[width=0.8\linewidth]{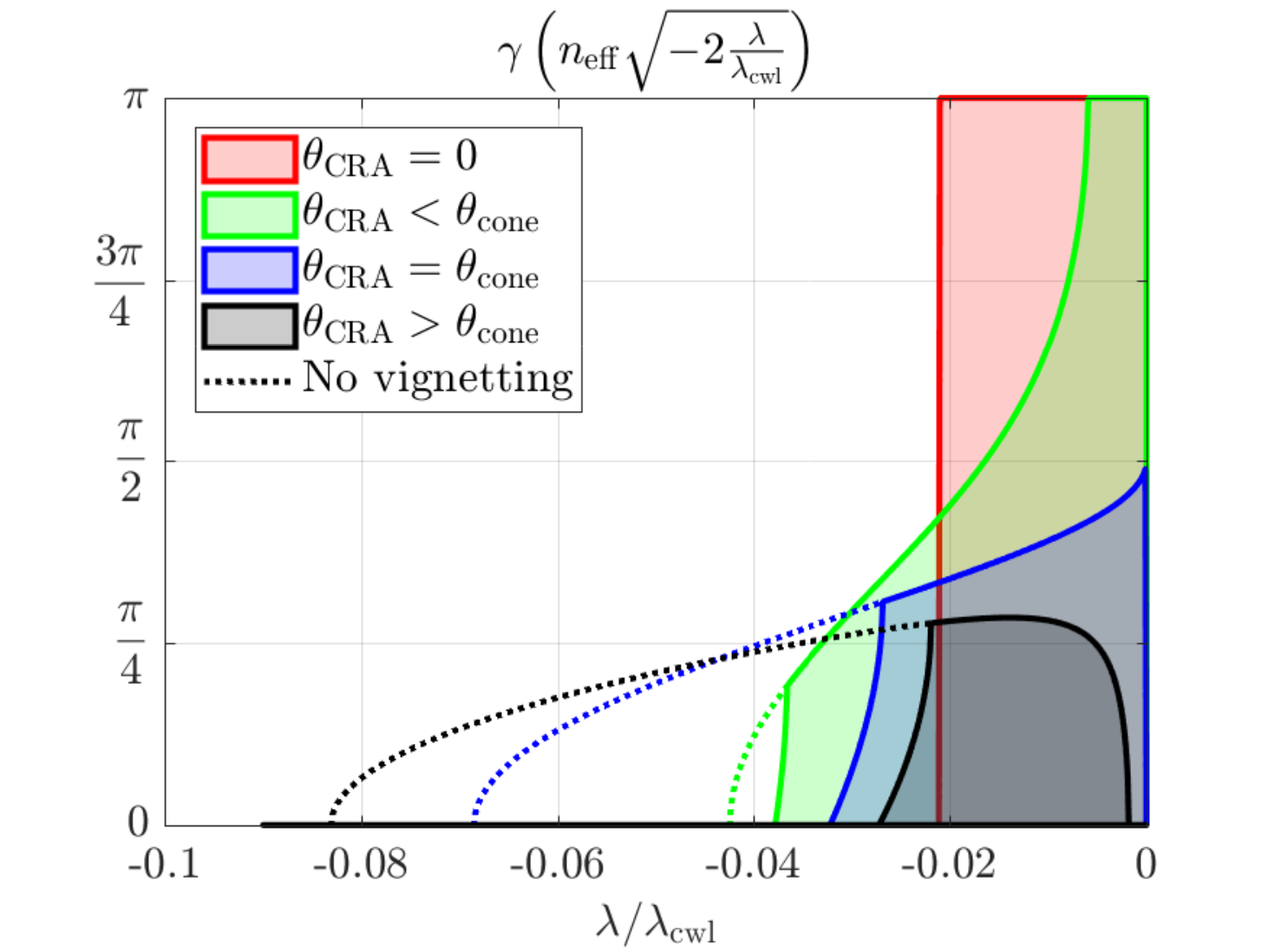}
		 \caption{\label{fig:gradual} The shape of the kernel is qualitatively shown for different regimes of $\cra$ and $\cone$.
		  Vignetting causes the kernel to be cut off. The dotted line continues the shape of the kernel in the absence of vignetting.}
\end{figure}

Initially, the kernel in \eqrefn{eq:asymptotic} is equivalent to the kernel in
the ideal finite aperture case (\eqrefn{eq:asymptoticold}). After the onset of
vignetting, the kernel is cut-off (Fig. \ref{fig:gradual}). Because
the kernel is less wide, the resulting transmittance will be shifted
less. In the next section we derive how to calculate this shift.

\subsection{Wavelength correction method}
\label{sec:org2dfd15d}
\label{org9f2472b}
Each filter has been designed to sense a specific wavelength. This
wavelength is defined as the central wavelength \(\lcwl\)  of the transmittance
when illuminated in orthogonal collimated conditions. To plot the
spectrum, the response for each filter is plotted at this central wavelength \(\lcwl\).

As discussed however, the actual central wavelength of the filter
depends on the angular distribution of the incident light. Thus, to correctly plot a spectrum, the central
wavelength needs to be corrected for each filter, taking its physical
position into account.

The shift in central wavelength can be quantified by the mean of the kernel (when interpreted as a
distribution) \cite{Goossens2018}. The mean is defined as
\begin{eqnarray}
\label{eq:expectedvalue}
\bar{\lambda}_{\cone,\,\cra}
      &=& \dfrac{\displaystyle \int_{\lambda_\text{min}}^{\lambda_\text{max}} \lambda  K_{\cone,\,\cra}(\lambda) \intd \lambda  }{\displaystyle \int_{\lambda_\text{min}}^{\lambda_\text{max}}  K_{\cone,\,\cra}(\lambda) \intd \lambda  } \nonumber\\
      &\stackrel{\tiny \tiny \text{(normalized)}}=& \displaystyle \int_{\lambda_\text{min}}^{\lambda_\text{max}} \lambda  K_{\cone,\,\cra}(\lambda) \intd \lambda.
\end{eqnarray}
In Appendix \ref{org9acc08e} we explain how to numerically approximate this integral.

To calculate the new central wavelength of the filter the mean value
\(\bar{\lambda}\) is added to the original central wavelength
such that
\begin{eqnarray}
\label{eq:cwlcorrect}
\lcwl^\text{new} &=& \lcwl + \bar{\lambda}_{\cone,\,\cra}.
\end{eqnarray}
The negative mean value is added because the shifts are towards
shorter wavelengths.

\subsection{Model parameter estimation}
\label{sec:org2c84704}
\label{org5b1dfa1}

In this section a practical algorithm is presented to estimate the
model parameters based on the vignetting profile.

The vignetting profile can be measured by
illuminating the entrance pupil uniformly with diffuse light from an integrating sphere. 

The vignetting profile does not only describe optical
vignetting. Natural vignetting, pixel vignetting and pupil aberrations
might also be present. An algorithm is therefore required to isolate the
contribution of optical and mechanical vignetting.

Optical vignetting is characterized by an abrupt change in the
vignetting profile \cite{Asada1996}. This abrupt change is caused by
the vignetting circle which suddenly starts cutting off part of the
exit pupil. This discontinuity can be exploited to isolate
optical vignetting from other contributions to the intensity fall-off.

The onset of optical vignetting is when the vignetting circle just
touches the border of the aperture. This happens when the the
vignetting circle has moved a distance equal to the difference in
radii such that
\begin{equation}
d_v= P-R.
\end{equation}
Rearranging gives 
\begin{equation}
\label{eq:onset}
P - d_v = P - h \tan \cra  = R = x \tan \cone.
\end{equation}

For each vignetting profile measured at a different f-number, an
equation of the form of \eqrefn{eq:onset} can be written down. The
corresponding chief ray angles are obtained from the vignetting
profile at the point where the abrupt change starts.

The unknown model parameters \(P\) and \(h\) can then be estimated from
solving the (possibly overdetermined) linear system

\begin{equation}
\label{eq:modelsystem}
\begin{bmatrix}
1 &-\tan(\theta_{\text{CRA},1}) \\
\vdots & \vdots \\
1 &-\tan(\theta_{\text{CRA},n}) \\
\end{bmatrix}
\begin{pmatrix}
P\\h
\end{pmatrix}
=
\begin{pmatrix}
R_1\\
\vdots\\
R_n
\end{pmatrix}.
\end{equation}

To determine the radius \(R\) of the exit pupil, the working f-number \(f_{\#,W}\)
must be used. The working f-number is calculated as (\cite{Kingslake1992})
\begin{equation}
\label{eq:workingfnumber} 
f_{\#,W} \approx \left(1+\frac{m}{m_P}\right) f_{\#},
\end{equation}
with \(m\) and \(m_P\) being the linear and pupil magnification
respectively.

The definition of the f-number then implies that
\begin{equation}
\label{eq:Rwork}
R = x\tan\cone = \dfrac{x}{2 f_{\#,W}}.
\end{equation}

\section{Experimental validation}
\label{sec:orgd445e24}

In this section, we estimate the model parameters and demonstrate that
using \eqrefn{eq:cwlcorrect} the shifted spectra in measurements can be corrected.

\subsection{Experimental setup}
\label{sec:orgd8bba6d}

We make use of Imec's VNIR Snapscan camera \cite{Pichette2017a}. In the Snapscan the image
sensor is placed behind the lens and its position can be controlled
using an automated translation stage. This makes the system ideally
suited for controlling the chief ray angles. The Snapscan is used with
an Edmund Optics 16 mm C Series VIS-NIR fixed focal length lens
because it has large chief ray angles and has considerable vignetting.

\begin{figure}[h] 
\centering
	 \begin{subfigure}{0.45\linewidth} 
	 \includegraphics[width=\linewidth]{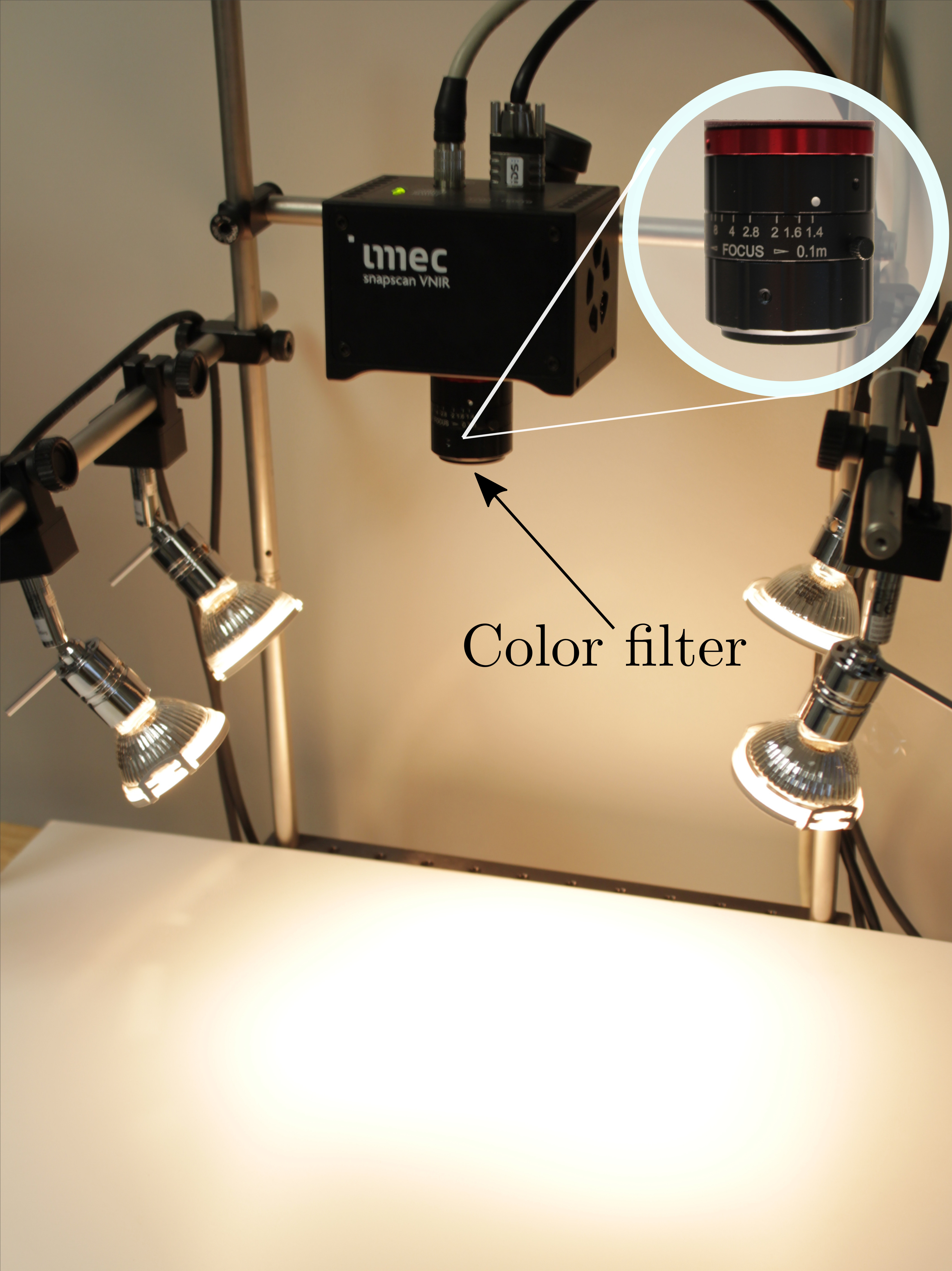}
	 \caption{\label{fig:visnirlens}Imec's Visible Near-Infrared Snapscan camera.} 
	 \end{subfigure}
	 \begin{subfigure}{0.45\linewidth} 
		 \includegraphics[width=\linewidth]{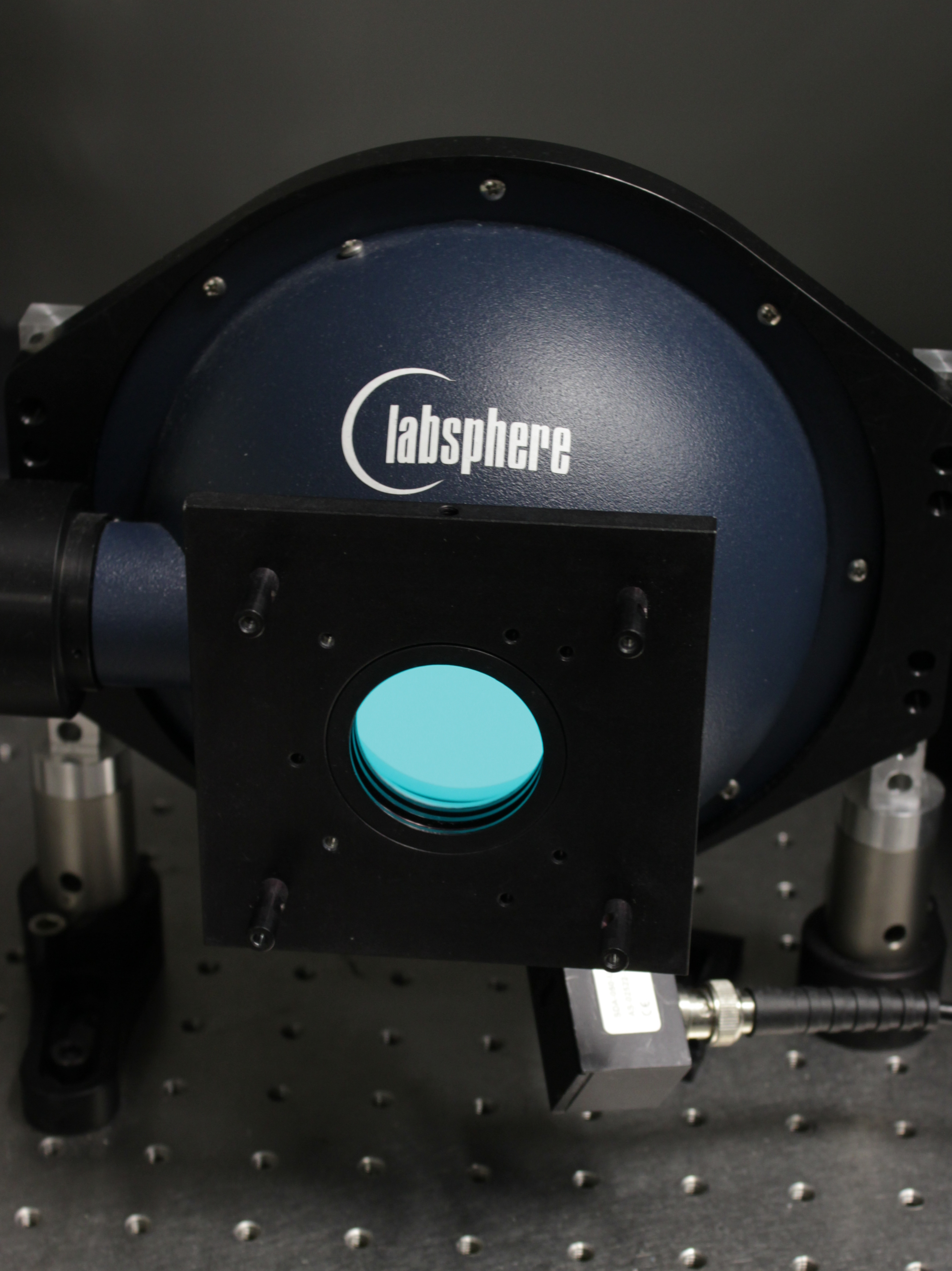}
		 \caption{Integrating sphere to measure vignetting profiles. \label{fig:sphere}} 
	\end{subfigure}
\caption{\textbf{Experimental setups.}\label{fig:setups}}
\end{figure}

To efficiently compare different points in the scene, a color filter
was placed in front of the lens. The scene was filled with a uniform
white reference tile (Fig. \ref{fig:visnirlens}). The
combination of the filter and white reference effectively creates a
scene-filling uniform spectral target. 

The transmittance of the color filter is calculated using a flat-fielding approach. This
requires taking three images: a dark image (no light) and two images
of the white reference tile, one with and one without color filter.
Using the notation of the spectral imaging model
(\eqrefn{eq:pixel-kernel}), the transmittance is then calculated as
\begin{equation}
\label{eq:transmittance}
\text{transmittance} = \dfrac{\text{DN}_\text{filter} - \text{DN}_\text{dark}}{\text{DN}_\text{white} - \text{DN}_\text{dark}}.
\end{equation}

The Edmund Optics 16 mm lens is focused at the target surface
(approx. 24 mm from the lens).

\subsection{Model parameter estimation}
\label{sec:org6708b89}

We apply the model parameter estimation method from \eqrefn{eq:modelsystem} to an Edmund Optics
16 mm lens. The exit pupil \(x=21\, \text{mm}\), linear magnification \(m = 0.06\) and pupil
magnification \(m_P = 1.3\). The filters have an effective refractive
index of \(\neff=1.7\).

\begin{figure}[H]
\centering
		 \includegraphics[width=0.6\linewidth]{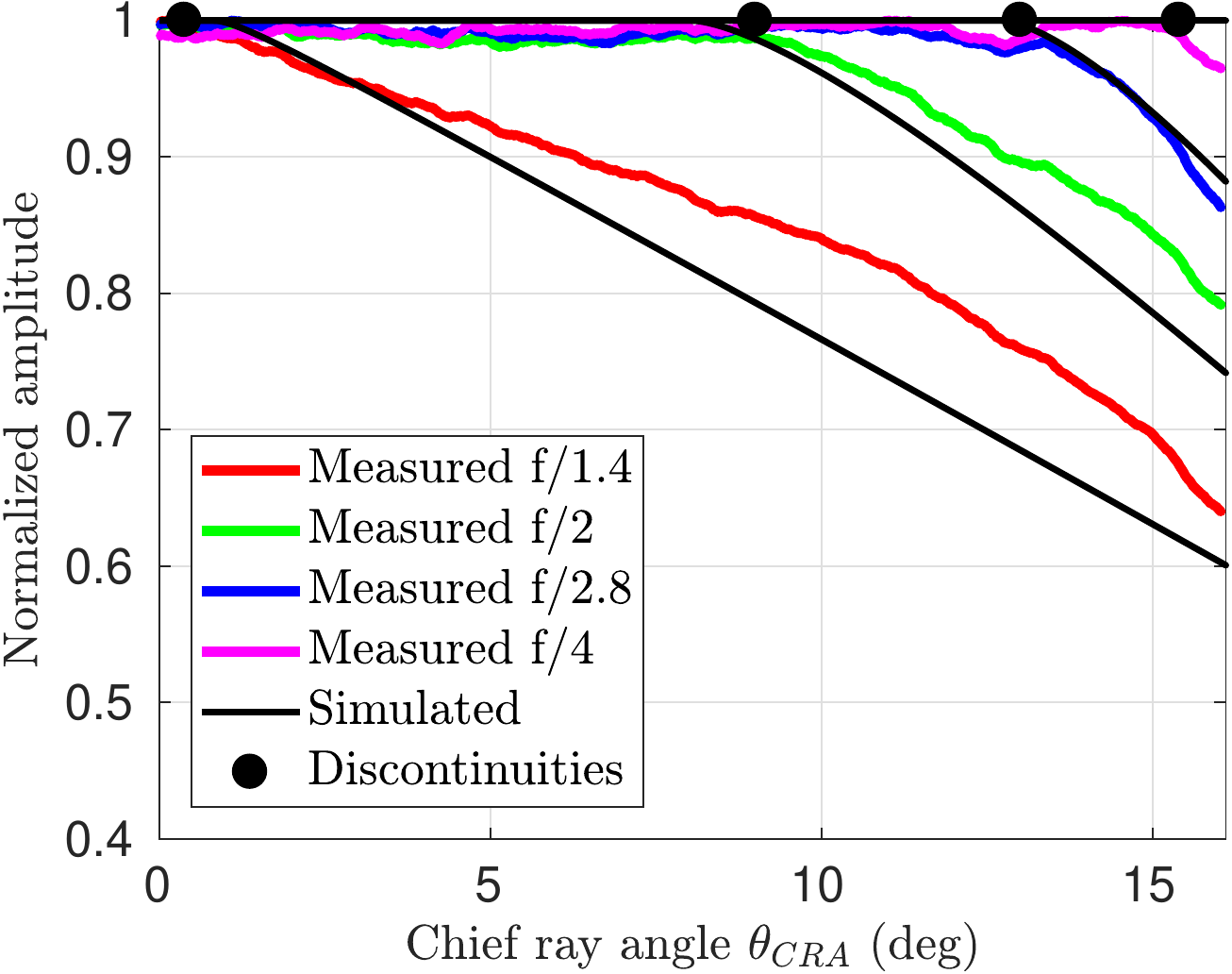}
		 \caption{\label{fig:eo16-profile} Vignetting profile for Edmund Optics 16 mm lens measured for different f-numbers. 
		 Using the estimated model parameters from \eqrefn{eq:estimated}, the vignetting profile is predicted (black line).}
\end{figure}

The vignetting fall-off profiles were measured using a camera with a
panchromatic image sensor. The camera was illuminated with uniform
light from an integrating sphere (Fig. \ref{fig:sphere}). In the
vignetting profiles there are several discontinuities, marked by black
dots in Fig. \ref{fig:eo16-profile}. To isolate the discontinuities in the 
vignetting profile we found that they are best visualized when the
vignetting profile is divided by the profile 
measured for a high f-number (Fig. \ref{fig:eo16-profile}). Since for
this high f-number there is no optical vignetting, this approach seems
to cancel out some of the other contributions.  This operation is
merely for visualization purposes and does not impact the fitting
procedure since the position of the discontinuities remain unchanged. 

From Fig. \ref{fig:eo16-profile} and \eqrefn{eq:modelsystem} it follows
that the following overdetermined system must be solved:

\begin{equation}
\label{eq:modelsystem-numbers}
\begin{bmatrix}
1 &-\tan(0.38^\circ) \\
1 &-\tan(9^\circ) \\
1 &-\tan(13^\circ) \\
1 &-\tan(15.4^\circ)
\end{bmatrix}
\begin{pmatrix}
P\\h
\end{pmatrix}	
=
\begin{pmatrix}
   7.1715\\
   5.0201\\
   3.5858\\
   2.5100
\end{pmatrix}.
\end{equation}
Where each angle corresponds to a black dot on
Fig. \ref{fig:eo16-profile}. And \(R_i\) was determined using \eqrefn{eq:Rwork}.

The solution of the overdetermined system, using the Moore–Penrose inverse, is
\begin{equation}
\label{eq:estimated}
 \begin{pmatrix}
 P\\h
 \end{pmatrix}
=
 \begin{pmatrix}
 7.4236	\\16.991
 \end{pmatrix}.
\end{equation}
We are in the \(h<x\) regime because \(16.99 < 21\) mm.

The deviations from the model prediction mean that there are other
effects that contribute to the vignetting profile. Nevertheless, the onset of
vignetting, as marked by the black dots, can be predicted very
well. The deviations might be due to the fact that natural vignetting changes with the aperture size \cite{Gardner1947} or because of
pupil aberrations. However, as we will see in the results, these
deviations have little impact on the correction of the shifted spectra.

\subsection{Experiment and data analysis}
\label{sec:orgfb01acc}

In each experiment, transmittance images (\eqrefn{eq:transmittance}) are taken at different
f-numbers using the setup shown in Fig. \ref{fig:visnirlens}. 
For each f-number, the spectrum is plotted at three chosen positions
that correspond to chief ray angles of \(1.9^\circ\), \(10.3^\circ\) and
\(17.4^\circ\).

To plot the spectra, the output (\eqrefn{eq:transmittance}) for each
pixel is plotted at the original central wavelength \(\lcwl\) of the deposited filter. Because of
this, the shifts of the filters towards shorter wavelengths creates
the illusion that the spectra move towards longer wavelengths. The
actual spectrum of the incident light, of course, does not change.

We plot three graphs. The first is the
uncorrected transmittance for all the samples. The second is the
transmittance corrected assuming an ideal finite aperture (\eqrefn{eq:idealcorrection}). The third
is the transmittance corrected using the vignetting model
(\eqrefn{eq:cwlcorrect}). 

For comparison, we also simulate the results
of the experiment. The simulation is done by combining
\eqrefn{eq:pixel-kernel} and \eqrefn{eq:transmittance}.

For each plot we also display two metrics that quantify the
improvement: the correlation coefficient and the maximum error. These
quantities were calculated for each pair of spectra. Only the worst
values are displayed on the graph.
\subsection{Results}
\label{sec:orgee4c31a}
\begin{figure}[H]
\centering
	 \begin{subfigure}{0.7\linewidth} 
		 \includegraphics[width=\linewidth]{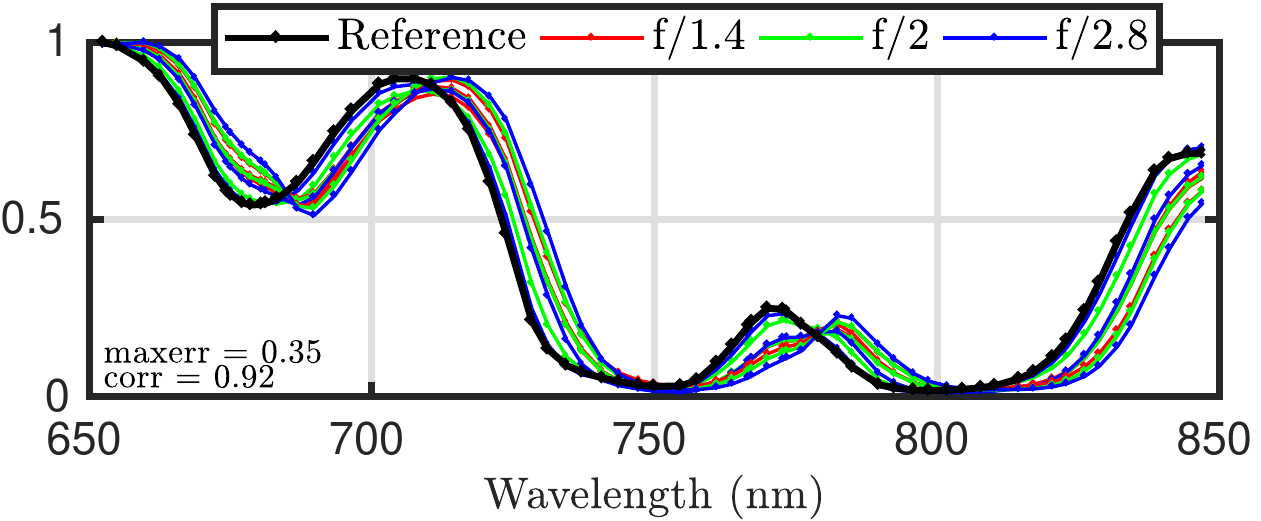}
		 \caption{Measured transmittance (uncorrected) \label{fig:eo16-fig1}}
	 \end{subfigure}
	 \begin{subfigure}{0.7\linewidth} 
		 \includegraphics[width=\linewidth]{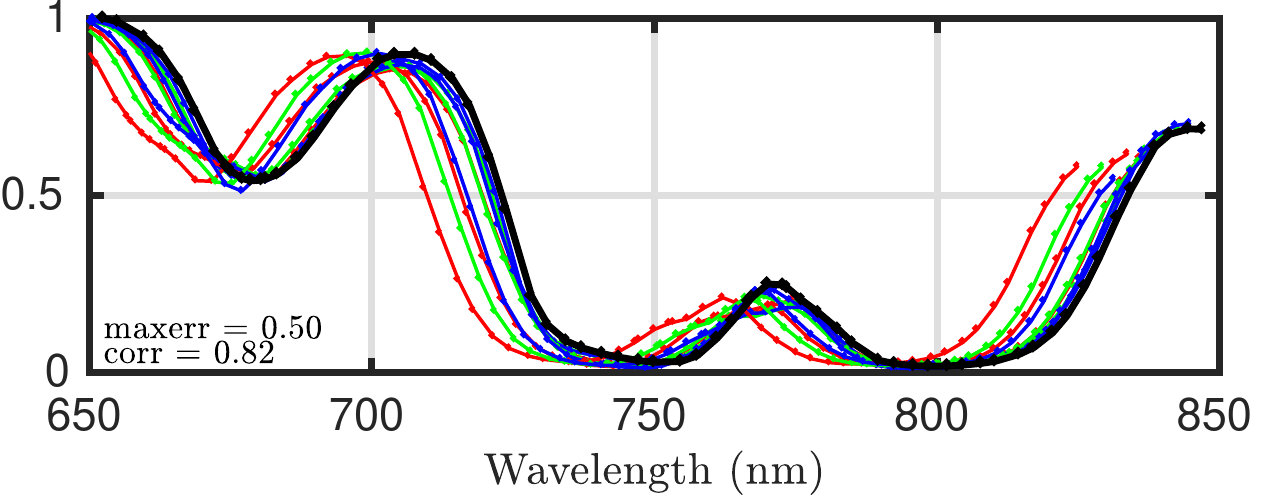}
		 \caption{Corrected using \eqrefn{eq:idealcorrection}\label{fig:eo16-fig2}} 
	 \end{subfigure}
	 \begin{subfigure}{0.7\linewidth} 
		 \includegraphics[width=\linewidth]{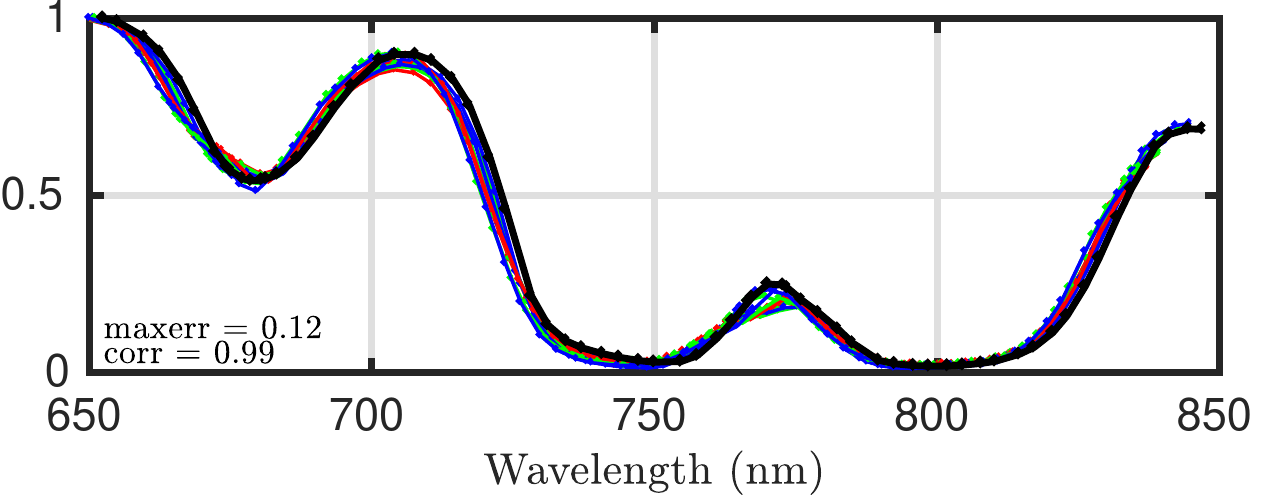}
		 \caption{Corrected using vignetting model \eqrefn{eq:cwlcorrect} \label{fig:eo16-fig3}}
	 \end{subfigure}     

		 \caption{\label{fig:eo16-craweep} \textbf{Edmund Optics 16 mm lens}. The uncorrected reflectance values demonstrate position and f-number dependent shifts. The spectrum is plotted at three chosen positions
that correspond to chief ray angles of $1.9^\circ$, $10.3^\circ$ and
$17.4^\circ$.  Without considering vignetting, the model overcorrects the shifts (Fig. \ref{fig:eo16-fig2}). Using \eqrefn{eq:cwlcorrect} with the estimated model parameters, the correction is significantly improved.} 
\end{figure}

\begin{figure}[H]
\centering
	 \begin{subfigure}{0.7\linewidth} 
		 \includegraphics[width=\linewidth]{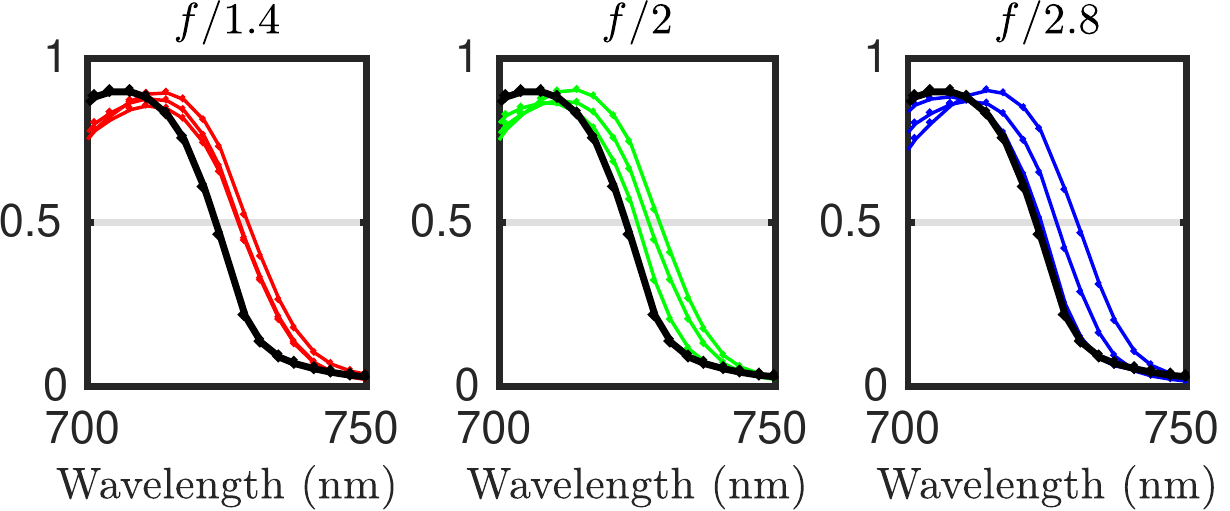}
		 \caption{Measured transmittance (uncorrected) \label{fig:eo16-spread-fig1}}
	 \end{subfigure}
	 \begin{subfigure}{0.7\linewidth} 
		 \includegraphics[width=\linewidth]{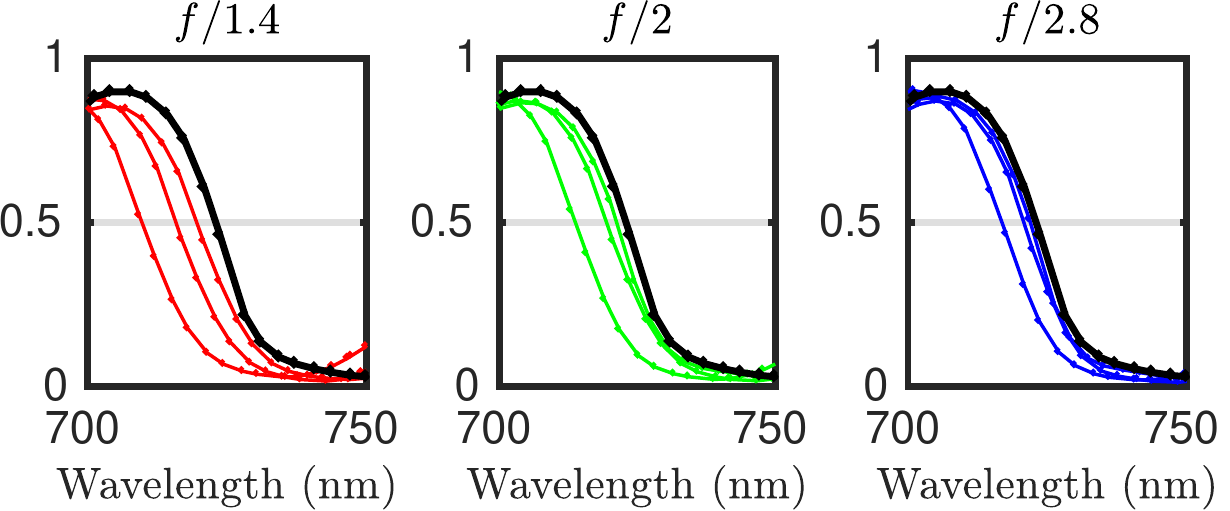}
		 \caption{Corrected using \eqrefn{eq:idealcorrection}\label{fig:eo16-spread-fig2}} 
	 \end{subfigure}
		 \caption{\label{fig:eo16-spread} Close-up of a region in Fig. \ref{fig:eo16-craweep} for each separate f-number. The spread of the spectra for $f/1.4$ is smaller than for larger f-numbers. 
		 The correction with \eqrefn{eq:cwlcorrect} is not shown here.}
\end{figure}

The uncorrected transmittances show clear
variations compared to the reference measured at \(f/8\)
(Fig. \ref{fig:eo16-fig1}). The variations are plotted for each
f-number separately in (Fig. \ref{fig:eo16-spread})

The different spectra measured at \(f/1.4\) are shifted less compared to
the higher f-numbers (Fig. \ref{fig:eo16-spread}). Intuitively one would expect larger
deviations for smaller f-numbers.
Without vignetting at \(f/1.4\), the spread between the
measured spectra for filters centered at 700 nm should be about 11
nm while the measured spread is about 2 nm.

Therefore, if one ignores the presence of optical vignetting and uses
\eqrefn{eq:idealcorrection} to correct the central wavelengths, the
shifts are significantly overcorrected (Fig. \ref{fig:eo16-fig2} and \ref{fig:eo16-spread-fig2}). 
Because optical vignetting is more pronounced for
smaller f-numbers, these are overcorrected the most.

If the vignetting model is used to correct the central wavelengths
using \eqrefn{eq:cwlcorrect}, the different spectra align very well, as
desired (Fig. \ref{fig:eo16-fig3}).

\begin{figure}[H]
\centering
	 \begin{subfigure}{0.7\linewidth} 
		 \includegraphics[width=\linewidth]{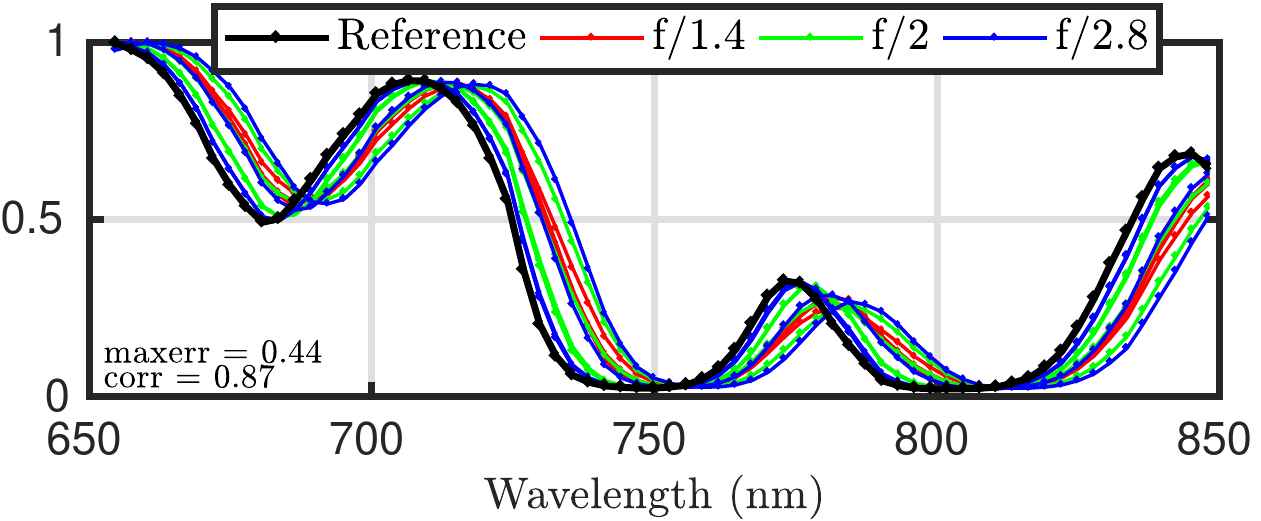}
		 \caption{Measured transmittance (uncorrected)}
	 \end{subfigure}
	 \begin{subfigure}{0.7\linewidth} 
		 \includegraphics[width=\linewidth]{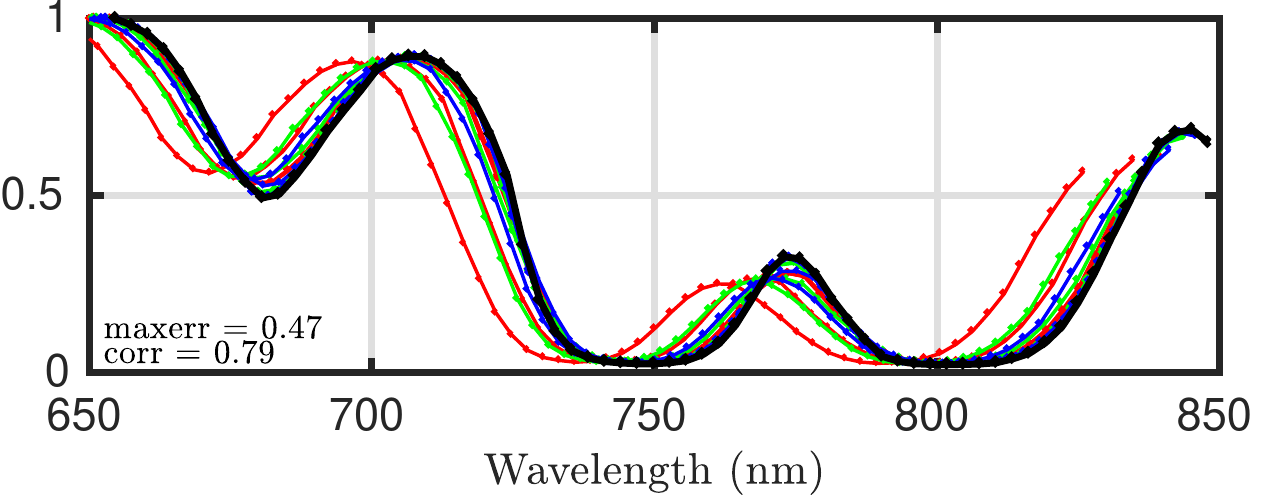}
		 \caption{Corrected using \eqrefn{eq:idealcorrection}} 
	 \end{subfigure}
	 \begin{subfigure}{0.7\linewidth} 
		 \includegraphics[width=\linewidth]{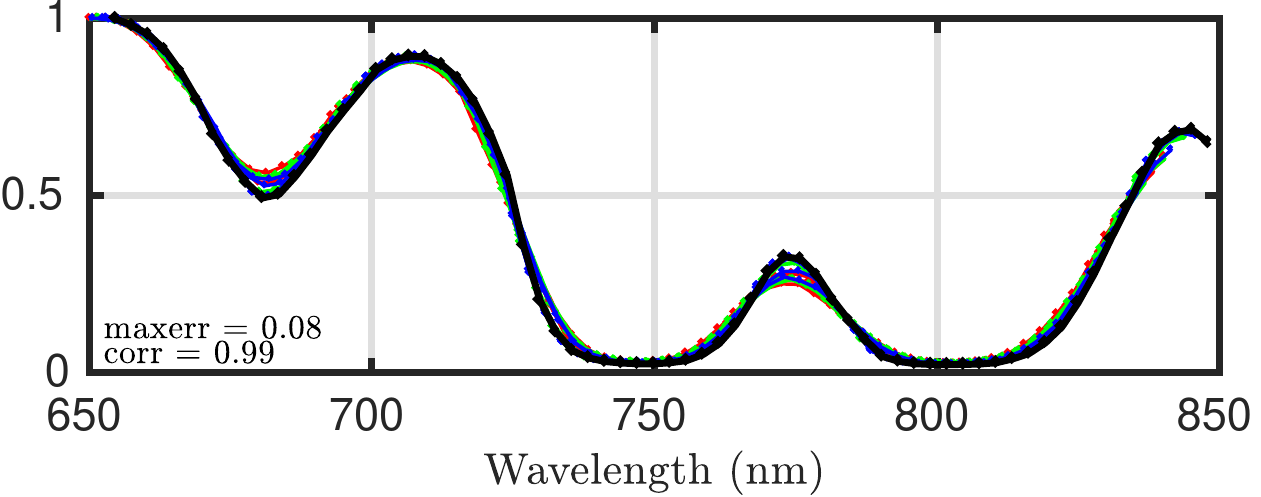}
		 \caption{Corrected using vignetting model \eqrefn{eq:cwlcorrect} \label{fig:sim-fig3}}
	 \end{subfigure}     
	 \caption{\label{fig:eo16-simulated-crasweep}\textbf{Simulated Edmund Optics 16 mm lens}. This is a simulation of the experimental conditions of Fig. \ref{fig:eo16-craweep}. The shifts in the uncorrected and corrected spectra (\eqrefn{eq:cwlcorrect}) are nearly identical to the real measurements.}
\end{figure}

In the simulation, very similar results are obtained
(Fig. \ref{fig:eo16-simulated-crasweep}). This shows that the model
can also be used to predict the impact of a lens with optical
vignetting on the measurements. It also shows that even in simulation
some variation remains after correction (Fig. \ref{fig:sim-fig3}). This is
because the kernel does not only shift the filters but also changes
their shape (Fig. \ref{fig:gradual}).

\begin{figure}[H]
\centering
		 \includegraphics[width=0.7\linewidth]{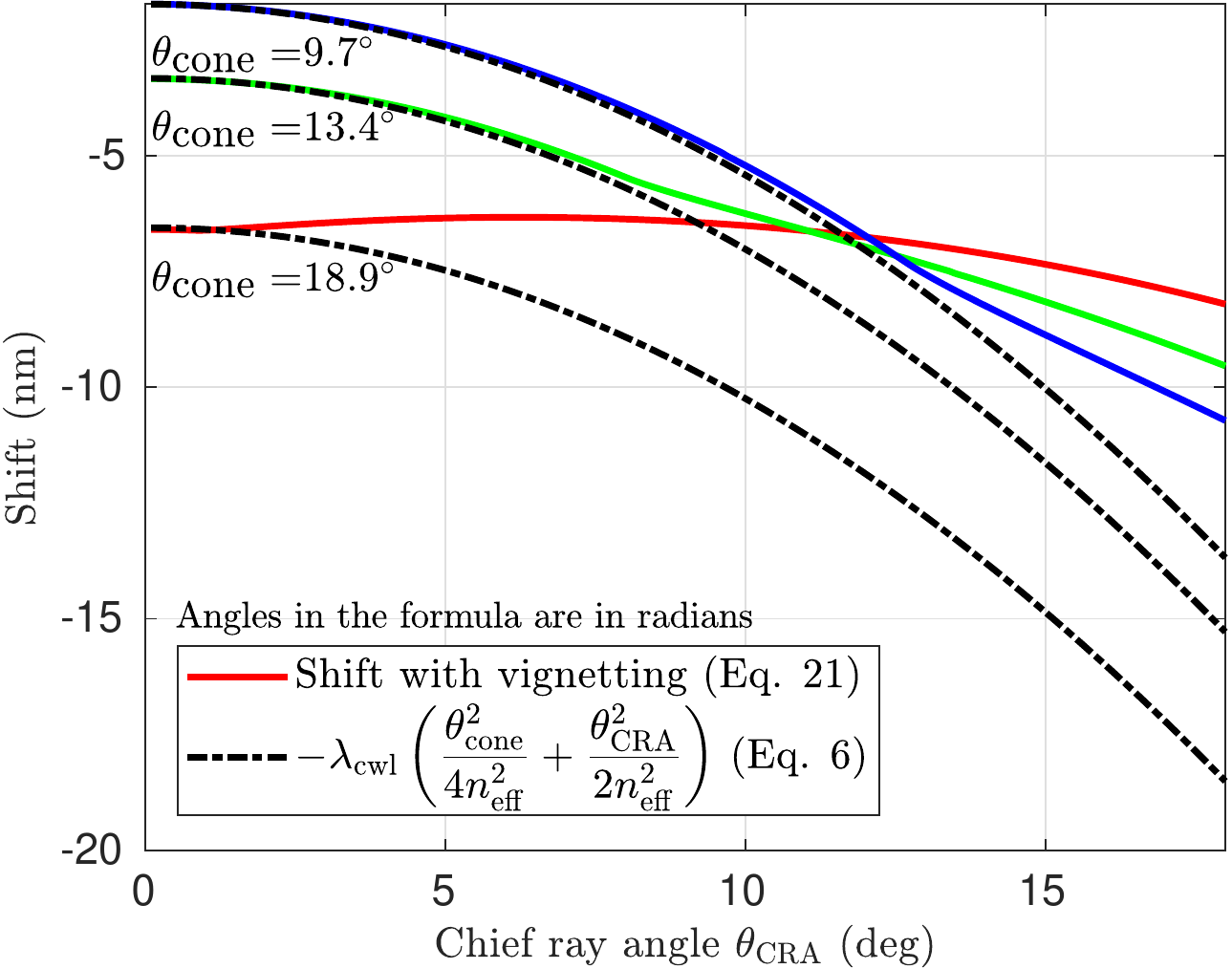}
		 \caption{\label{fig:shiftprofile}Numerical approximation of the mean value compared to the asymptotic approximation \eqrefn{eq:idealmean}.  The asymptotic approximation remains valid until the onset of optical vignetting. Simulated for $\lcwl = 700$ nm at f-numbers $f/1.4$ (red), $f/2$ (green) and $f/2.8$ (blue).}
\end{figure}

The differences between Fig. \ref{fig:eo16-fig2} and
Fig. \ref{fig:eo16-fig3} can be better understood by comparing  
\(\eqrefn{eq:idealmean}\) and \(\eqrefn{eq:expectedvalue}\) visually
(Fig. \ref{fig:shiftprofile}). This figure summarizes the effect of
vignetting on the shift of the filters. It shows that after the onset of vignetting,
\eqrefn{eq:idealmean} starts to overestimate the shift. The near
constant value of the red curve also explains why the spread in shift
for \(f/1.4\) is only about 2 nm.

\section{Discussion}
\label{sec:orgaff4845}

The results demonstrate that the proposed method can be used to
correct undesired shifts in measurements in the presence of optical
or mechanical vignetting.

Correcting these position-dependent shifts is important for improving
the performance of machine learning applications. And because vignetting
is a common phenomenon, our method will be relevant for many practical applications.

The method is powerful since it can treat a lens as a black-box. This
is important since often a complete physical model of the lens is
unavailable. But even when a full model is available, it is still
unpractical to get an idea of the dominant effects and how to correct
for them. Our method offers a very practical way to isolate the effect of
optical vignetting and estimate the model parameters to predict and correct it.

To make the effect of optical vignetting more understandable, more
analytical studies on the subject are needed. Currently, the
mean of the kernel is calculated numerically. An analytical
approximation of the mean in the vignetting regime would be useful for
intuitive understanding and quick calculations.

The vignetting model presented in this paper successfully generalizes the model
from prior work \cite{Goossens2018}. However, the model also has some potential
limitations which are discussed below.

First, there is the assumption that the vignetting circle moves
proportional to \(\tan\cra\) (\eqrefn{eq:dv}). This might not be a good
approximation for some lenses. The equation however can be easily
substituted without major changes to the correction method. 

Second, pupil aberrations might cause changes in the radii of the exit
pupil and projected vignetting circle. Pupil aberrations will be the topic of future work.

Third, theoretically there can multiple projected vignetting circles simultaneously present in the
system. This means that the aperture is cut off in more complex
ways. One example is pixel vignetting, which also
limits the cone of light that can reach the photodiode. This effect
was not studied in this article. However, in
Fig. \ref{fig:eo16-profile}, for f/1.4, there is arguably a second 
discontinuity around \(\cra\approx 11^\circ\). The extension to multiple
vignetting circles could be investigated further. However, in our
experiments the additional gains would be negligible.

\section{Conclusion}
\label{sec:org20c0b07}

In many applications it is important that the measured spectra are
independent of the lens and the position of the target in the
scene. 

Shifts in the measured spectra are caused by the sensitivity of the filters to the
angle of incidence and must be corrected for. In previous work we demonstrated
this for real lenses that exhibited no vignetting.

Vignetting, however, is a common phenomenon in real lenses and must be taken
into account for accurate correction. Thus, in this article we generalized
our model for a vignetted aperture. We demonstrated that this
generalization is vital for correcting the observed shifts in measured
spectra. 

Because vignetting is so common, our method has the potential to improve the performance of many spectral imaging
applications. These will mostly be applications that use fast lenses (large
aperture) or low cost lenses but still require high spectral accuracy.

\appendix

\section*{Appendix}

\section{Closed-form solution of the kernel}
\label{sec:org052096f}
 \label{org3f6ca83}
In this section we derive a closed-form solution of the kernel and
derive the analytical approximation given in \eqrefn{eq:asymptotic}.

The derivation of the closed-form solution is identical to derivation
presented in prior work (see \cite{Goossens2018}) up to a few minor
modifications.
First, the shape of the exit pupil changes and therefore also the
distribution of incident angles. 
Thus \(\gamma\) is defined by \eqrefn{eq:gamma} instead of  \(\gamma(\phi) = \eta(\phi)\).
Second, the area of the vignetted exit pupil changes. Therefore the kernel
should be normalized accordingly.
Therefore
\begin{equation}
\label{eq:kernelaftersubstitution}
\begin{array}{lcl}
	 K_{\cone,\,\cra}(\lambda) &=&  \dfrac{\displaystyle \int_{\phi_\text{min}}^{\phi_\text{max}}   2x^2 \gamma(\phi) \dfrac{\tan \phi}{\cos^2 \phi} \delta(\lambda - \Delta(\phi)) \intd \phi}{\displaystyle \iint\limits_{\tiny \substack{\text{Vignetted} \\ \text{exit pupil}}} \intd A}, \\
	 & = & 	 \dfrac{\displaystyle \int_{\phi_\text{min}}^{\phi_\text{max}}   2x^2 \gamma(\phi) \dfrac{\tan \phi}{\cos^2 \phi} \delta(\lambda - \Delta(\phi)) \intd \phi}{A(\cone,\cra)},\\
\end{array}
\end{equation}
with \(\Delta(\phi)\) defined in \eqrefn{eq:tiltshift} and
\(A(\cone,\cra)\) being the area of the vignetted exit pupil which is
defined as \cite{Asada1996}
\begin{equation}
A(\cone,\cra) = P^2 (\alpha-\sin(\alpha) \cos(\alpha))+R^2(\beta-\sin(\beta)\cos(\beta)),
\end{equation}
with  
\begin{eqnarray}
\alpha &=& \text{Re}\left(\arccos \dfrac{P^2-R^2+d_v^2}{2Pd_v}\right)\\
\beta &=& \text{Re}\left(\arccos\dfrac{R^2-P^2+d_v^2}{2Rd_v}\right),
\end{eqnarray}
and \(d_v = h\tan \cra\) and \(R= x\tan \cone\).

The integral in \eqrefn{eq:kernelaftersubstitution} can be rewritten
using the Heaviside function \(H(\cdot)\). If we define \(\Pi(\phi) =
H(\phi - \phi_\text{min}) - H(\phi - \phi_\text{max})\), then 
\begin{equation}
\label{eq:inftyintegral}
\begin{array}{lcl}
	 K_{\cone,\,\cra}(\lambda) & = & \dfrac{\displaystyle \int_{-\infty}^{\infty}   2x^2 \gamma(\phi) \dfrac{\tan \phi}{\cos^2 \phi} \delta(\lambda - \Delta(\phi)) \Pi(\phi) \intd \phi}{A(\cone,\cra)}.
\end{array}
\end{equation}

To proceed we use the same steps as described in Appendix A of
\cite{Goossens2018}, the only difference being the difference in sign
convention for shift. The closed-form analytical solution to \eqrefn{eq:inftyintegral} can be formulated as 
\begin{equation}
\label{eq:fullkernel}
 K_{\cone,\cra}(\lambda) = \dfrac{g(\lambda) \gamma(\Delta^{-1}(\lambda)) }{A(\cone,\cra)} x^2 \Pi\left(\Delta^{-1}(\lambda)\right),
\end{equation}
with \(\gamma\) as defined in \eqrefn{eq:gamma} and the function
\(g(\lambda)\) containing all remaining terms:
\begin{eqnarray}
\label{eq:g}
g(\lambda) &=& \dfrac{2\neff^2}{\lcwl}\cdot\dfrac{\left(1+\dfrac{\lambda}{\lcwl}\right)}{\left(1 + 2\neff^2\dfrac{\lambda}{\lcwl} + \neff^2\dfrac{\lambda^2}{\lcwl^2}\right)^2}\\
	   &=& \frac{2\,\neff^2}{\lcwl} + \mathcal{O}\left(\dfrac{\lambda}{\lcwl}\right),\quad \text{for } \dfrac{\lambda}{\lcwl} \rightarrow 0.
\end{eqnarray}
The notation \(\mathcal{O}(\cdot)\) is the Big O notation which describes
the limiting behavior of the error term of the approximation.

Combining the approximation in \eqrefn{eq:g} with
\begin{equation}
\begin{array}{l}
\Delta^{-1}(\lambda) = \arcsin\left(\neff \sqrt{-\dfrac{\lambda}{\lcwl} \left(2+\dfrac{\lambda}{\lcwl} \right)} \right)\\
= \neff \sqrt{-\dfrac{2\lambda}{\lcwl}} + \mathcal{O}\left(\dfrac{\lambda^{3/2}}{\lcwl^{3/2}}\right),\quad \text{for } \dfrac{\lambda}{\lcwl}\rightarrow 0,
\end{array}
\end{equation}
the kernel can be approximated as
\begin{equation}
\label{eq:appendix-approx}
K_{\cone,\,\cra}(\lambda) \sim \dfrac{2\,\neff^2}{\lcwl} \cdot \dfrac{\displaystyle \gamma\left(\phiapproxmin \right)}{A(\cone,\cra)} x^2.
\end{equation}
The only difference with the result from earlier work (\eqrefn{eq:asymptotic}) is the value of
\(\gamma\) and that the area of the vignetted exit pupil now also varies with \(\cra\).

\section{Calculating the mean of the kernel}
\label{sec:org471a331}
\label{org9acc08e}

The mean \(\bar{\lambda}\) is required in the central wavelength correction
formula of \eqrefn{eq:cwlcorrect}. The mean is defined by the integral
\begin{equation}
\displaystyle \int_{\lambda_\text{min}}^{\lambda_\text{max}} \lambda  K_{\cone,\,\cra}(\lambda) \intd \lambda.
\end{equation}

In this article we numerically approximated the mean value and used this value for correction.
For this, the limits \(\lambda_\text{min}\) and
\(\lambda_\text{max}\) of the integral need to be known. By
construction \(\gamma(\phi)\) becomes zero where needed (\eqrefn{eq:gamma}). Because of this and
the fact the shift is always less than or equal to zero, the upper limit can
always be taken zero in numerical integration. Therefore \(\lambda_\text{max} = 0\).

The lower integration limit however should be known exactly since
there is no theoretical limit to its negative value. 

For each case, there are two options. The maximal radius is
constrained by either the exit pupil or the projected vignetting circle.
If it is constrained by the exit pupil, \(r_\text{max}=R+d\). If it
is constrained by the vignetting circle, the result depends on
whether \(h\geq x\) or \(h<x\).

In the case that \(h>x\), the kernel becomes zero if \(\eta = \nu\).
The maximal radius if found by equating \eqrefn{eq:eta} and \eqrefn{eq:nu} and
solve to r such that
\begin{equation}
r = \pm \frac{\sqrt{(d-d_r)[d(P^2-d_r^2) + d_r
(d^2-R^2)]}}{d-d_r},
\end{equation}
of which only the positive solution is meaningful.
For \(h<x\), the maximal radius is \(r=P+d_r\). 

The maximal radius \(r_\text{max}\) for all cases can be formulated compactly as 

\begin{equation}
 r_\text{max} = \begin{cases} 
      \min(R+d,P+d_r) & h<x  \\
      \min\left(R+d,\frac{\sqrt{(d-d_r)[d(P^2-d_r^2) + d_r (d^2-R^2)]}}{d-d_r}\right)& h\geq x
   \end{cases}.
\end{equation}
The min function models that either the exit pupil or the
projected vignetting circle is the limiting circle.

From \(r_\text{max}\), the maximal incident angle \(\phi_\text{max}\) can
be calculated. By definition, \(\phi_\text{max}  = \arctan
\dfrac{r_\text{max}}{x}\). 
And finally, using \eqrefn{eq:tiltshift}, 
\begin{equation}
\lambda_\text{min} = \Delta(\phi_\text{max}).
\end{equation}
Using the chosen integration limits, the solution is then
approximated using a left Riemann sum. A step size = 0.01 nm was used.

However for small values of \(\cone\), the kernel will converge to a
Dirac distribution. Therefore a much smaller step size might be
required. Instead it is advised to use the asymptotic approximation
\eqrefn{eq:idealmean}. In the presence of optical vignetting the
asymptotic approximation is valid for \(d_v < P-R\). Or alternatively,
\begin{equation}
\label{eq:cravalid}
\cra < \arctan \dfrac{P-R}{h}.
\end{equation}
This condition specifies the onset of
vignetting, after which the numerical solution should be used. The numerical solution and the asymptotic approximation
are both plotted in Fig. \ref{fig:shiftprofile}.

This method is implemented in Code 1. A usage example is given in Code 2.

\section*{Acknowledgements}
We thank Nick Spooren for proofreading the manuscript.

\bibliographystyle{abbrv}
\bibliography{/home/thomas/Documents/imec/phd/library/library.bib}
\end{document}